\documentclass[a4paper,fleqn,usenatbib]{mnras}

\usepackage[T1]{fontenc}
\usepackage{ae,aecompl}

\usepackage{graphicx}	
\usepackage{amsmath}	
\usepackage{amssymb}	

\newcommand{\re}{R$_{\rm e}$}

\newcommand{\simlt}{\lower.5ex\hbox{$\; \buildrel < \over \sim \;$}}
\newcommand{\simgt}{\lower.5ex\hbox{$\; \buildrel > \over \sim \;$}}

\newcommand{\nad}{$\rm NaD$}

\newcommand{\cfs}{$\rm C4668$}

\newcommand{\mgfep}{$\rm [MgFe]'$}
\newcommand{\mgfe}{$\rm [Mg/Fe]$}

\newcommand{\nafe}{$\rm [Na/Fe]$}

\newcommand{\cfe}{$\rm [C/Fe]$}
\newcommand{\ofe}{$\rm [O/Fe]$}

\newcommand{\fet}{$\rm Fe3$}

\newcommand{\hbo}{$\rm H\beta_o$}
\newcommand{\hb}{$\rm H\beta$}

\newcommand{\mgb}{$\rm Mgb5177$}

\newcommand{\kms}{\,km\,s$^{-1}$}
\newcommand{\afe}{$[\rm \alpha/{\rm Fe}]$}
\newcommand{\afep}{$\rm [Z_{Mg}/Z_{Fe}]$}

\newcommand{\gammab}{$\rm \Gamma_b$}

\newcommand{\zh}{$\rm [M/H]$}

\newcommand{\amiles}{$\rm \alpha \! - \! MILES$}

\newcommand{\nai}{$\rm NaD$}
\newcommand{\naii}{$\rm NaI8190$}
\newcommand{\naiii}{$\rm NaI1.14$}
\newcommand{\naiv}{$\rm NaI2.21$}
\newcommand{\alj}{$\rm \alpha_{Na_j}$}

\newcommand{\gaj}{$\rm \gamma_j$}
\newcommand{\Teff}{$\rm T_{eff}$}
\newcommand{\Feh}{$\rm [Fe/H]$}
\newcommand{\Msol}{$\rm M_\odot$}
\newcommand{\alfa}{$\rm \alpha$}

\title[Na features in Early-type Galaxies]{IMF and \nafe\ abundance ratios from optical and NIR Spectral Features in Early-type Galaxies}

\author[F. La Barbera et al.]
{F. La Barbera$^{1}$\thanks{E-mail:  labarber@na.astro.it},
A., Vazdekis$^{2,3}$, I. Ferreras$^{4}$, A. Pasquali$^{5}$, 
C. Allende Prieto$^{2,3}$, \and 
B. R\"ock$^{2,3}$, D.~S. Aguado$^{2,3}$, R.~F. Peletier$^{6}$ \\
$^{1}$INAF-Osservatorio Astronomico di Capodimonte, sal. Moiariello
16, Napoli, 80131, Italy\\
$^{2}$Instituto de Astrof\'\i sica de Canarias, Calle V\'\i a L\'actea s/n, E-38205
  La Laguna, Tenerife, Spain\\
$^{3}$Departamento de Astrof\'\i sica, Universidad de La Laguna (ULL), E-38206  La Laguna, Tenerife, Spain\\
$^{4}$Mullard Space Science Laboratory, University College London, Holmbury St Mary,
  Dorking, Surrey RH5 6NT, UK\\
$^{5}$Astronomisches Rechen-Institut, Zentrum f\"ur Astronomie, Universit\"at Heidelberg, 
M\"onchhofstr. 12-14, D-69120 Heidelberg, Germany\\
$^{6}$ Kapteyn Astronomical Institute, University of Groningen, P. O. Box 800, 9700 AV Groningen Netherlands\\
  }

\date{Submitted to MNRAS (revised version \today)}

\pubyear{2016}

\begin{document}
\label{firstpage}
\pagerange{\pageref{firstpage}--\pageref{lastpage}}
\maketitle

\begin{abstract}
We present a joint analysis of the four most prominent
sodium-sensitive features { 
(Na \! D, Na \! I $\lambda$8190\AA, Na \! I $\lambda$1.14$\mu$m, and Na \! I $\lambda$2.21$\mu$m)},
in the
optical and Near-Infrared spectral range, of two nearby, massive
($\sigma \sim 300\,$\kms ), early-type galaxies (named XSG1 and XSG2).
Our analysis relies on deep VLT/X-Shooter long-slit spectra, along
with newly developed stellar population models, allowing for
\nafe\ variations, up to $\sim$1.2\,dex, over a wide range of age,
total metallicity, and IMF slope.  The new models show that the
response of the Na-dependent spectral indices to \nafe\ is stronger
when the IMF is bottom heavier.  For the first time, we are able to
match all four Na features in the central regions of massive
early-type galaxies, finding an overabundance of \nafe, in the range
0.5--0.7\,dex, {\it and} a bottom-heavy IMF.  Therefore, individual
abundance variations cannot be fully responsible for the trends of
gravity-sensitive indices, strengthening the case towards a
non-universal IMF.  Given current limitations of theoretical
atmosphere models, our \nafe\ estimates should be taken as upper
limits.  For XSG1, where line strengths are measured out to
$\sim$0.8\re, the radial trend of \nafe\ is similar to \afe\ and \cfe,
being constant out to $\sim$0.5\re, and decreasing by
$\sim$0.2--0.3\,dex at $\sim$0.8\re, without any clear correlation
with local metallicity.  Such a result seems to be in contrast with
the predicted increase of Na nucleosynthetic yields from AGB stars and
Type II SNe.  For XSG1, the Na-inferred IMF radial profile is
consistent, within the errors, with that derived from TiO features and
the Wing-Ford band, presented in a recent paper.
\end{abstract}




\section{Introduction}
\label{Sec:Intro}

Although constraining the initial mass function (hereafter IMF) of
unresolved stellar populations poses a number of difficult challenges,
it has been possible to discern a significant variation of the low-mass
end of the IMF in early-type galaxies (ETGs). Constraints on the IMF have been
achieved on two fronts, namely via estimates of the stellar M/L through
dynamical \citep[see, e.g.,][]{Capp:12,Capp:13} or lensing analyses
\citep[see, e.g.,][]{Treu:10,Leier:15}; or
via targeted spectral features sensitive to the presence of low-mass
stars, which in effect constrain the relative contribution of dwarf
versus giant stars \citep[see, e.g.,][]{Cenarro:2003,vdC:10,Ferr:13,LB:13, Spiniello:2014}. These
two approaches are affected by different systematics, but both agree
in that the stellar populations found in the more massive ETGs appear
to have formed from a (bottom-)heavier IMF than the standard one \citep[see,
e.g.,][]{Kroupa:01,Chabrier:03}. 
The more homogeneous
populations typically found in massive ETGs simplify -- to some degree
-- the analysis of the strong degeneracies found in the interpretation
of gravity-sensitive spectral features. However, some issues of the
analysis are not fully understood. These issues are mainly related to
the sensitivity of the indices to non-solar abundance ratios, both as
a general trend between the Fe-group and alpha-group elements (measured
as [Mg/Fe] or [$\alpha$/Fe]), or as more specific variations in {\sl
individual} elements -- described here as [X/Fe] -- that affect the
specific gravity-sensitive features under scrutiny.

The steps ahead in the interpretation of gravity-sensitive line
strengths as a signature of a non-universal IMF involve a painstaking,
detailed analysis of the inherent degeneracies of these
indices. Although age and metallicity can be robustly taken into
account by use of additional spectral indices, or by spectral fitting,
the effect of [$\alpha$/Fe] or, even more so, [X/Fe] is much harder to
quantify, due to the lack of a fully consistent, extensive, library of
stellar spectra (empirical or theoretical) covering a wide range of
the parameter space. An empirically motivated approach is always
preferred -- given the complexity of the line strengths and the
stellar populations under study. For instance, in \citet{LB:13} it was
possible to model the [$\alpha$/Fe] dependence of the indices by use
of stacks of SDSS spectra selected at fixed velocity dispersion and
[$\alpha$/Fe]. Such an approach allowed us to discard, in an empirical
way, the possibility that the trends found in gravity-sensitive
indices were solely caused by variations in [$\alpha$/Fe]. However, at
the moment, it is not possible to base the analysis entirely on an
empirical approach, and one has to resort to more uncertain stellar
atmosphere calculations \citep[see, e.g.,][]{TMJ11,CvD12a}.

More empirical information is needed to advance in this field, and to
improve theoretical models of stellar atmospheres. On this front, we
turned towards a detailed analysis of the spectra of individual ETGs
at very high S/N, covering a wide spectral window.  In this context,
\citet{LB:16} presented a high quality spectrum of a massive ETG taken
by the VLT/X-Shooter spectrograph, to explore the local (i.e. radial)
variations in the IMF by use of TiO gravity-sensitive indices in the
optical and NIR, as well as the Wing-Ford (FeH-dependent) band in the
NIR. A combination of three TiO indices allowed us to break the
degeneracies of individual line strengths, leading to a robust
determination of a radial gradient in the IMF slope, being
bottom-heavy in the central region, approaching the standard case at
around an effective radius. This result is consistent with previous
studies \citep{NMN:15a}. Surprisingly, the Wing-Ford index was not as
strong as expected, and its radial variation was rather mild,
apparently suggestive of no IMF change.  However, one should note that
each gravity-sensitive index depends on different intervals of the IMF
stellar mass range. The Wing-Ford band is especially sensitive to the
very low-mass end (i.e. the feature is only prominent in the coolest
M-dwarfs). Therefore, this result is consistent with a scenario where
the IMF of massive ETGs has an excess of low-mass stars (around a
spectral type M0V, corresponding to a stellar mass 0.5\,M$_\odot$) but
not of very low-mass stars (approximately below a spectral type M4V,
corresponding to 0.2\,M$_\odot$), i.e. a tapering of the IMF
functional form, similar to that for the Milky-Way IMF. In this case,
a single power law (also described as a unimodal IMF), is readily
ruled out \citep{LB:16}. Alternatively, one may consider that the
Wing-Ford band is more strongly anticorrelated with \afe\ than the
theoretical predictions, hampering its joint interpretation with the
optical TiO indices.

The ultimate goal of the interpretation of gravity-sensitive spectral
features is to determine the local property that drives the variations
in the IMF.  Various theoretical analyses propose an excess of
low-mass stars through turbulent fragmentation in the environment of a
strongly starbursting interstellar medium \citep[see,
  e.g.][]{PadNor:02,Hopkins:13,Chabrier:14}. On first principles one
could expect a significant variation in the IMF between a relatively
quiet environment and a more extreme scenario -- found in the cores of
massive ETGs -- where a substantial mass in stars ($\sim
10^{11}$M$_\odot$) forms within a relatively short period of time
(1-2\,Gyr) and within a small volume (r$<$1-2\,kpc). Moreover, in
order to match the observed properties of massive ETGs (most notably,
their metallicity and $\alpha$-enhancement), a time-dependent scenario
is required, where a top-heavy phase, expected during the first stages
of galaxy evolution, is followed by a bottom-heavy
phase~\citep{Vazdekis1996, Vazdekis1997, Weidner2013, Ferreras2015}.
A deeper understanding of the information encoded in gravity-sensitive
spectral indices will allow us to determine the major driver of IMF
variations.  The large metallicity radial gradient found in (some) massive
ETGs~\citep[see, e.g.][]{LB:12} is indicative of at least two
different components: a massive, compact core, formed early, and an
extended envelope, possibly made up of accreted material from
mergers. This two-stage formation paradigm \citep[e.g.,][]{Oser:10,
  NavarroGonzalez:2013} should also lead to two different stellar mass
functions. Indeed, such a variation in the IMF was found in recent
studies based on long-slit and IFU spectroscopy
\citep{NMN:15a,NMN:15b}. The lack of an IMF radial gradient in a
relic, compact, massive galaxy \citep{NMN:15c} was encouraging within
this simple model. Furthermore, these spatially resolved studies
allowed us to focus on the {\sl local} drivers of IMF
variations. Although the first results pointed to a significant
correlation with central velocity dispersion
\citep[][]{Ferr:13,LB:13, Spiniello:2014, LB:15, NMN:15a}, the CALIFA observations of a
set of ETGs revealed that metallicity, rather than {\sl local} velocity
dispersion, might be the primary driver of IMF
variations~\citep{NMN:15b}. It is worth emphasizing here that the
SDSS-based work is fully consistent with this interpretation, as a
significant correlation is expected between metallicity and the
fibre-integrated velocity dispersion measured in the SDSS
spectra. However, the CALIFA results were mostly based on a single
gravity-sensitive feature (TiO2), and more work is doubtlessly needed
to disentangle the underlying degeneracies.

As a next step, we turn our focus towards the sodium-based indices.
One of the key observables originally proposed to explore the low-mass
end of the IMF in quiescent galaxies is the NIR Na {\sc I} feature at
8,190\AA\ \citep{SpinTa:71}. The strength of this feature in massive
ETGs could be interpreted either as an excess of low-mass stars with
respect to the Milky Way standard IMF; as a product of the expected
super-solar [$\alpha$/Fe]; or even as an overabundance in the
individual elements that affect this index, most notably [Na/Fe]. A
combination of indices, including the NaD feature in the optical,
might allow us, in principle, to disentangle the effect of abundance
variations from the presence of low-mass
stars~\citep{Spiniello:2012}. However, such an analysis has been
rather inconclusive so far, especially in light of recent results
targeting specific (one, or two; mostly optical) Na-sensitive
features.  \citet{SmithLucey2013} found a giant elliptical galaxy with
a lightweight IMF normalization -- as constrained by strong lensing --
in apparent contrast with its { Na \! I $\lambda$8190\AA} line, as strong as in other
high-$\sigma$ ellipticals. \citet{smith:2015} also concluded that
state-of-the-art stellar population models struggle to reproduce the NIR
\naiii\ feature in most massive galaxies without breaking other
constraints.  \citet{MenesesGoytia2015} also found that both their
models and those of~\citet{CvD12a} are unable to reproduce the
observed large { Na \! I $\lambda$2.21$\mu$m values} of massive ETGs ({ see also R\"ock et al.,
submitted}).  Recently, \citet{Ziel:15} found a highly Na-enhanced
population in the bulge of M31 -- along with a Na-depleted population
in M32, still pointing to difficulties of the most recent stellar
population models to match both the \nai\ and \naii\ absorption
lines. Furthermore, \citet{McConnell:15} suggested large gradients in
[Na/Fe] in the central regions of ETGs.  These findings show that an
in-depth analysis of sodium-dependent spectral features, in both the
optical and NIR, is very much needed.

In this work, we perform, for the first time, a joint, radially
extended, analysis of the four Na-sensitive spectral features, 
{ Na \! D, Na \! I $\lambda$8190\AA, Na \! I $\lambda$1.14$\mu$m, and Na \! I $\lambda$2.21$\mu$m
(hereafter  \nai , \naii , \naiii , and \naiv, respectively)
}
in two
massive ($\sigma\sim$300\,\kms ) ETGs at redshift $z\sim$0.06.  The
two targets, hereafter XSG1 and XSG2, have been selected from the
SPIDER survey~\citep{SpiderI}, more specifically from the subsample of
160 ETGs with 280$<\sigma<$320\,\kms\ defined by \citet[][hereafter
  LB13]{LB:13}.  For both galaxies, we have gathered new, very deep,
``high'' resolution (R$>$4000), long-slit spectroscopy with the
X-Shooter spectrograph at the ESO-Very Large Telescope (VLT).  Our
main goal is to constrain, simultaneously, the effect of abundance
ratios on Na lines (i.e. \afe\ and \nafe ), as well as the slope of
the stellar IMF in these galaxies. To model the effect of \nafe , we
rely on new stellar population models (named Na--MILES) developed specifically for the
purpose of this work, extending previous models, covering the optical
and NIR spectral range, from~\citet{Vazdekis:12}
and~\citet{RV:15}. The effect of \afe\ is modelled with an empirical
approach. The results presented in this work are intended to provide a calibration
for the interpretation of Na-sensitive features in ETGs. To this effect,
our newly developed  Na-enhanced stellar population models are made publicly 
available.

The layout of the paper is as follows. In Sec.~\ref{sec:data}, we
describe the data. The new stellar population models, and the
measurement of Na line strengths, are described in
Sec.~\ref{sec:models} and Sec.~\ref{sec:indices}, respectively.
Sec.~\ref{sec:sppars} deals with the determination of stellar
population parameters, such as \cfe\ and \afe , that are relevant for
our analysis.  The fitting methodology leading to the constraints on
\nafe, IMF slope and the sensitivity of the Na indices to \nafe\ is
presented in Sec.~\ref{sec:fitting}. Some important caveats are
discussed in Sec.~\ref{sec:upperlimits}. The results are presented in
Sec.~\ref{sec:results}, followed by a discussion and a summary in
Secs.~\ref{sec:disc} and~\ref{sec:summary}, respectively. We include two appendices with detailed
information about the effect of low T$_{\rm eff}$ stars on the
Na-MILES models (App.~\ref{app:teff}) and on a comparison between
CvD12 and our models (App.~\ref{app:rescaled}).

\section{Data}
\label{sec:data}

The targets of the present work are XSG1 
(SDSS J142940.63+002159.0) and XSG2 (SDSS J002819.30-001446.7),
two massive (M$_\star\!\sim$1--2$\times10^{11}$M$_\odot$;
$\sigma\sim$300\kms) ETGs at redshift z$\sim$0.05574 and
z$\sim$0.06009, respectively. Both galaxies are selected from the SPIDER
survey~\citep{SpiderI}, more specifically from the subsample of $160$
ETGs with $280<\sigma<320\,$\kms\ defined in LB13.

XSG1 has spherical morphology (b/a$\sim$0.9, corresponding to an E0/E1
type), whose light distribution is well described by a de Vaucouleurs
model, with an effective radius of \re$\sim\,4.1^{\prime\prime}$ (see
LB16 for details).  XSG2 has a flatter light distribution, with
b/a$\sim$0.67.  Using the software 2DPHOT~\citep{LBdC08} to fit the
SDSS r-band image of XSG2 with a two-dimensional, PSF-convolved,
S\'ersic model (as for XSG1, see LB16), gives a best-fit S\'ersic
n$\sim$5.4, and an effective radius of \re$\sim$5.4$^{\prime\prime}$.
\citet{Simard:11} report a total \re$\sim$4.2$^{\prime\prime}$ from the galaxy
bulge+disk decomposition, while from SDSS, we retrieve a value of
\re$\sim$4.1$^{\prime\prime}$, more consistent with the total \re . Throughout the
present work~\footnote{The value of \re\ is only used to
  scale the galacto-centric distance of spectra extracted in different
  radial bins. Hence, the adopted value of \re\ does not affect at all
  our conclusions.}, for XSG2, we adopt the SDSS value of
\re$\sim$4.1$^{\prime\prime}$ (i.e. the same value as for XSG1).

For both targets, we have obtained new, deep long-slit spectroscopy
with the X-Shooter spectrograph at the ESO-VLT, on Cerro Paranal
(Proposal IDs: 092.B-0378, 094.B-0747; PI: FLB).  X-Shooter is a
second-generation ESO-VLT instrument -- a slit echelle spectrograph
that covers a wide spectral range (3000--25000\,\AA), at relatively
high resolution~\citep{Vernet:2011}. The { incoming beam} is split into
three independent arms (ultraviolet-blue, UVB: 3000--5900\,\AA;
visible, VIS: 5300--10200\,\AA; near-infrared, NIR: 9800--25000\,\AA),
each one equipped with optimized optics, dispersive elements, and
detectors, ensuring high sensitivity throughout the entire spectral
range. Data for XSG1 (XSG2) were taken through five (ten) observing
blocks (OBs), each including two exposures on target, interspersed by
two (one) sky exposures, with the same integration time as for the
science target.  This setup resulted into a total on-target exposure
time of $\sim$1.7, 1.9, and $2.1$\,hr, in the X-Shooter UVB, VIS, and
NIR arms, respectively. For each arm, the data were pre-reduced using
version 2.4.0 of the data-reduction pipeline~\citep{Mod:2010},
performing the subsequent reduction steps (i.e. flux calibration, sky
subtraction, and telluric correction) with dedicated FORTRAN software
developed by the authors. We refer the reader to LB16 for a detailed
description of each reduction step (see also~\citealt{SCH:2014}).

For the present study, we use the six radially binned spectra of XSG1
analyzed by LB16 (see their figure~4), obtained by folding up data
from opposite sides of the X-Shooter slit around the photometric
centre of the galaxy.  In order to minimize seeing effects, the
innermost spectrum has a width of 1.3$^{\prime\prime}$ (i.e. $\pm
0.675^{\prime\prime}$) around the photometric centre of the galaxy,
corresponding to a factor of 1.5 times the mean seeing FWHM of the
data. The bin size is increased adaptively outwards, in order to
ensure a median S/N$>$90 per \AA\ in the optical spectral range (from
4800 to 5600\,\AA).  The outermost bin for XSG1 reaches an average
galacto-centric distance of $\sim$0.8\,\re .  For XSG2, because of the
lower~\footnote{The lower S/N ratio of the X-Shooter spectrum of XSG2
  is due to the fainter mean surface brightness of the galaxy (with
  respect to XSG1), rather than a lower data quality. } S/N, we did
not extract spectra at different radii, producing only the central
galaxy spectrum, extracted within an aperture of $\pm
0.675^{\prime\prime}$ (i.e. $\sim$0.2\re) around the galaxy
photometric centre, i.e.  the same aperture as for the innermost bin
of XSG1.  Thus, the present analysis is based on a total of seven
radially binned spectra, 6 for XSG1 plus 1 for XSG2.  The median S/N
measured in the central bins reaches 320 and 220 per \AA\ for XSG1
(see table~1 of LB16) and XSG2, respectively.  The main properties of
the seven binned spectra, including their galacto-centric distance
R/\re\ and velocity dispersion ($\sigma$), are summarized in
Tab.~\ref{tab:abundances}. The spectra of XSG1 are displayed,  over 
the whole  X-Shooter spectral range, in figure~4 of LB16. For XSG1, the $\sigma$ is measured as
detailed in LB16, running the software {\sc pPXF}~\citep{Cap:2004} on
different spectral regions of the UVB and VIS X-Shooter arms
($\lambda\lambda =4000-9000$\,\AA ), combining the corresponding
probability distribution functions into final estimates. For the
central spectrum of XSG2, the same approach gives
$\sigma\!\sim\!308\pm 3$\,\kms, in fair agreement with the SDSS-DR7
estimate ($292\pm 11\,$\kms), measured within the (larger) SDSS
1.5$^{\prime\prime}$-radius fibre aperture.

\section{Stellar population models}
\label{sec:models}

\subsection{Extended--MILES models}

We use the extended--MILES stellar population models, covering the
spectral range from $0.35$ to $5 \mu$m, at moderately high
resolution. Such wide wavelength baseline is achieved by combining
single-age, single-metallicity, stellar population (SSP) spectra
computed in various spectral ranges with our population synthesis
code, in a self-consistent manner. The optical SSP spectra
(\citealt{MILESIII}; recently updated by \citealt{Vazdekis:15},
hereafter V15) are extended out to $\sim0.9\mu$m, forming the
so-called MIUSCAT SSPs~\citep{Vazdekis:12}, and joined to those at
redder wavelengths~\citep{RV:15}, as described in \citet{RV:16}. All
these models are extensively based on empirical stellar libraries,
namely MILES in the optical range \citep{MILESI}, which together with
Indo-US \citep{Valdes04} and CaT \citep{CATI} are employed in the
MIUSCAT models, and the IRTF stellar library \citep{IRTFI,IRTFII}
reaching out to 5\,$\mu$m. The atmospheric parameters (\Teff, $\log
g$, \Feh) of stars in the different libraries, have been homogenized
to match the calibration of MILES \citep{MILESII}.

The models use two sets of scaled-solar theoretical isochrones, namely
the ones of \citet{Pietrinferni04} (BaSTI) and \citet{Padova00}
(Padova00). In the case of BaSTI, non-canonical models are used, with
the mass loss efficiency of the Reimers law \citep{Reimers} set to
$\eta=0.4$. The BaSTI isochrones are supplemented with the stellar
models of \citet{Cassisi00}, which allow us to cover the very low-mass
(VLM) regime down to $0.1$\,\Msol. The temperatures of these low-mass
stars are cooler than those of Padova00 \citep{Vazdekis:12}. Although
the Padova00 isochrones are defined over a wide range of
metallicities, the BaSTI models, supplemented by additional
computations as described in V15, allow us to virtually double the
number of metallicity bins. These two sets of isochrones include the
thermally pulsing AGB regime using simple synthetic prescriptions,
providing a significantly smaller contribution for this evolutionary
phase at intermediate-aged stellar populations in comparison to the
models of \citet{Marigo08} and \citet{Maraston05}. Finally, there are
differences between the BaSTI and Padova00 isochrones, described in
detail in \citet{Cassisietal04}, \citet{Pietrinferni04} and V15.
Throughout this paper, in order to allow for a direct comparison
with results from our previous works (e.g.~F13, LB13, and LB16), we
rely on models based on the Padova00 isochrones.

To transform theoretical isochrone parameters into stellar fluxes, we
use relations between colours and stellar parameters from empirical
photometric libraries. We mainly use the metallicity-dependent
relations of \citet*{Alonso96} and \citet*{Alonso99} for dwarfs and
giants, respectively. We also use the metal-dependent bolometric
corrections of \citet*{Alonso95} and \citet{Alonso99}, for dwarfs and
giants. We refer the interested reader to V15 for a complete
description of the method and relevant references.

The synthetic spectrum of an SSP is obtained by integrating along the
isochrone, following the approach described in V15, for the ``base''
models (see below). The number of stars within a mass bin is given by
the adopted IMF. The spectrum of a star -- given by a choice of of
atmospheric parameters (\Teff, $\log g$, \Feh) -- is obtained by
applying a local interpolation scheme that takes into account the
local density of reference spectra from the stellar library in this
three-dimensional parametric space. The procedure is detailed in
\citet{CATIV}, and updated in V15 (see their appendix~B). The interpolation
scheme is optimized to overcome the limitations imposed by gaps and
asymmetries in the distribution of reference stars in parameter
space.

Models are constructed with four different IMF shapes, as described in
\citet{CATIV} and V15. We use the multi-part power-law IMFs of
\citet{Kroupa:01}, i.e. universal and revised, and two power-law IMFs,
as defined in \citet{Vazdekis1996}, i.e unimodal (single-segment) and
bimodal. The lower and upper mass-cutoff is set to $0.1$ and
$100$\,M$_\odot$, respectively. The unimodal and bimodal IMFs are
defined by their logarithmic slope, $\Gamma$ and $\Gamma_b$,
respectively. { For reference, the \citet{Salpeter55} IMF is obtained when
adopting a unimodal IMF with $\Gamma=1.35$, whereas the Kroupa
Universal IMF is closely approximated by a bimodal IMF with
$\Gamma_b=1.3$.
Since the bimodal distribution consists of a power-law
at the high-mass end -- smoothly tapered towards low masses -- varying
$\Gamma_b$ changes the dwarf-to-giant ratio in the IMF through the
normalization. 
While this approach is different with respect to a
change  of the IMF slopes at low- and very-low mass (e.g.~CvD12b), this parameterisation is 
good enough for our purposes,  as in the present work we do not aim at
constraining the IMF shape in detail, but rather to perform a comparison of observed
and model Na line-strengths, adopting an IMF functional form that is consistent with 
constraints from our previous work (LB16).}

The (empirical) stellar libraries employed in our models unavoidably
follow the characteristic abundance pattern imprinted by the SFH of
our Galaxy. Since we use scaled-solar isochrones, and do not take into
account the specific elemental abundance ratios of the stellar
spectra, for which we simply assume $\rm [M/H]=$\Feh, these models can
be regarded as ``base'' models (see V15). Base models are
approximately self-consistent, and scaled-solar at around solar
metallicity.  In contrast, at low metallicity they lack consistency
since we combine scaled-solar isochrones with \alfa-enhanced stellar
spectra. In V15 we computed self-consistent models in the optical
range, that are either scaled-solar or \alfa-enhanced, at all
metallicities, with the aid of theoretical stellar spectra from
\citet{Coelho05,Coelho07}. Hereafter, we refer to such models as
$\alpha$--MILES\footnote{The synthetic spectra of $\alpha$--MILES
  models only cover the optical range ($\lambda\lesssim$7400\AA).}.
Briefly, to obtain an \alfa-enhanced model in the high metallicity regime,
we compute a reference scaled-solar SSP spectrum using MILES and
\alfa-enhanced isochrones \citep[from][]{Pietrinferni06}, along with
the corresponding scaled-solar and \alfa-enhanced SSP
spectrum based on the theoretical models of \citet{Coelho05}.
The residual between these two SSP spectra defines the
differential response due to the varying abundance ratio. Therefore, we
apply this correction to the reference SSP MILES-based scaled-solar
spectrum to obtain the desired \alfa-enhanced data. For the purpose of
the present work, we only use $\alpha$--MILES SSPs  to compare the
{\sl empirically derived} constraints of $\Delta$\nai/$\Delta$\afe\ to
model predictions (see Sec.~\ref{sec:aferesponses}). In general, the
analysis of observed line strengths of ETGs is based on
extended--MILES and Na--MILES (see \S\ref{sec:namiles}) model predictions.

\subsection{Na--MILES models}
\label{sec:namiles}
In this paper, we need to account for the effect of varying
\nafe. Therefore, we modify the modelling approach followed in our
previous papers. The most relevant differences with respect to V15
(where we compute $\alpha$-, rather than { Na-enhanced } models) are
that {\it i)} we fully rely on scaled-solar isochrones, and {\it ii)}
we apply theoretical differential corrections to \nafe\ overabundance,
specifically computed for each individual stellar spectrum in our
empirical library. Notice that, in contrast, the \alfa-enhanced models
were based on differential corrections for {\sl SSP spectra}.
Furthermore, our \nafe-enhanced models are not restricted to the MILES
spectral range, as we extend them out to the K-band. Hereafter, we
refer to these models as Na--MILES models.

Following \citet{Meszaros:12}, we create synthetic stellar spectra
for each of the 180 stars (from the IRTF library) used to build up our
models.  This choice is motivated by the fact that the stellar
parameter coverage of this library is dense enough at around solar
metallicity, leading to SSP models with good quality \citep[see
  figures~5 and 8 in ][]{RV:15}.  We obtained model atmospheres by
interpolation of the ODFNEW\footnote{\tt
  http://kurucz.harvard.edu/grids.html} ATLAS9 models of
\citet{CastelliKurucz04}, using {\it kmod}\footnote{\tt
  http://hebe.as.utexas.edu/stools}.  The ODFNEW models are available
for stars with effective temperatures higher than $3500$\,K. For
cooler stars in the IRTF library we adopt a fixed temperature of
$3500$\,K. The models adopt a reference solar composition from
\citet{GrevesseSauval98}.  
Notice that the main difference between the solar
abundances of \citet{GrevesseSauval98} and more recent versions (e.g.
\citealt{Asplund09}) are related to carbon, nitrogen, and oxygen
abundances, which { are up to 0.2\,dex higher} than in the previous
version. We scale the metal abundances according to \Feh\ derived for
spectra in the IRTF library, and use \alfa-enhanced models (by
$0.4$~dex) for stars with $\Feh<-1$.

The spectral synthesis was performed with the ASS$\epsilon$T code
\citep{Koesterke08,Koesterke09}, with line opacities compiled by
Kurucz, with updates to atomic damping constants published by
\citet{Barklem00}, including transitions of H$_2$,CH, C$_2$, CN, CO,
NH, OH, MgH, { SiH, SiO, and TiO}.  The equation of state included the first
92 elements in the periodic table and 338 molecules
(\citealt{Tsuji73}; with some updates), with partition
functions from \citet{Irwin81}. 
{ 
We did not include  $\rm H_2O$ transitions in the line opacity calculations, since these become 
important for effective temperatures lower than about 3000~K, 
i.e. below our threshold of 3500~K (see above).}
We accounted for bound-free absorption
from H, H$^{-}$, ,He{\sc I} and He{\sc II}, and the first two
ionization stages of C, N, O, Na, Mg, Al, Si, and Ca from the Opacity
Project \citep[see, e.g.][]{Cunto93} and Fe from the Iron Project
\citep{Nahar95,Bautista97}.  All models are computed under the
assumption of Local Thermodynamical Equilibrium or LTE.  We refer the
reader to \citet{Allende-Prieto08} for more details.

We compute synthetic stellar spectra for a range of \nafe, from 0 to
$+$1.2\,dex in steps of 0.3\,dex, covering the spectral range from
$0.35$ to $2.5\mu$m. { We stress that model atmospheres were not computed for
each  [Na/Fe] abundance ratio, but were interpolated from the \citet{CastelliKurucz04}
model grid, and therefore assuming a solar [Na/Fe] abundance ratio. The effect of 
variations in the Na abundance on the atmospheric structure is expected to be modest, 
since this element produces very limited opacity and only contributes free electrons 
in the outermost photospheric layers of the coolest stars~\citep{Meszaros:12},  
and therefore it was only considered at the spectral synthesis stage.
}
The models are computed for a total metallicity
\zh$>-0.25$~dex, i.e. within the values found in massive ETGs. For a
given value of \nafe, we divide each Na--enhanced, theoretical,
spectrum by its scaled-solar counterpart, to obtain the
(multiplicative) differential response to \nafe , which is applied to
the empirical stellar spectrum. This procedure is applied to all the
stars in the IRTF library. The correction for the other stellar
libraries (MILES, Indo-US, CaT), applies our local interpolation
algorithm to obtain the corresponding Na responses. Note that for each
library the relevant parameters of the interpolator are adjusted to
account for the maximum density of stars in that library
\citep[see][]{CATIV}. We smooth the theoretical stellar spectra,
adjusting the effective resolution to match the instrumental configurations of
the different stellar libraries. The spectral resolution is kept
constant with wavelength (at FWHM=2.5\AA), for all libraries, except
for the IRTF set, which is characterized by a constant
$\sigma=$60\,\kms.

\section{Line strengths}
\label{sec:indices}

In order to account for the effect of stellar population properties
(i.e.  age, metallicity, and abundance ratios) on Na-sensitive
features, we measure for each spectrum a set of spectral indices: 
\mgb; \fet$\rm =(Fe4383+Fe5270+Fe5335)/3$;
\cfs\ \citep[see][]{Trager98}; the total metallicity indicator
\mgfep~\citep{TMB:03}; as well as \hbo\
\citep[the optimized \hb\ index defined
by][]{CV09}.

\begin{table*}
\centering
\small
 \caption{Main properties of the spectra used in the present work.
   Cols.~1, 2 report the galaxy name, and the galacto-centric distance
   in units of \re\ (see Sec.~\ref{sec:data}), respectively.
   Columns~3, 4, 5, 6, 7 give velocity dispersion, age, total
   metallicity, \afe , and \cfe\ abundance ratios, respectively.
   Uncertainties are quoted at the 1$\sigma$ level, and represent
   internal (statistical) errors.  Notice that age and metallicity are
   estimated through the fitting of Na lines and \mgfep\ (see
   Sec.~\ref{sec:fitting}), while \afe\ and \cfe\ are derived as
   described in Sec.~\ref{sec:sppars}.  }
  \begin{tabular}{c|c|c|c|c|c|c}
  \hline
 Galaxy & $\rm R/R_e$ & $\sigma$ &  Age  & \zh  & \afe & \cfe \\
        &             &  (\kms ) & (Gyr) &  dex & dex  &  dex \\
   (1)  &     (2)     &     (3)  &   (4) &  (5) &  (6) &  (7) \\
 \hline
XSG1 & $0$    & $333 \pm 3$ &  $10.3 \pm  0.8$  &  $0.38 \pm 0.02  $ & $ 0.38 \pm 0.03$ & $ 0.12 \pm 0.02$ \\
XSG1 & $0.19$ & $316 \pm 3$ &  $10.3 \pm  0.8$  &  $0.27 \pm 0.03  $ & $ 0.38 \pm 0.04$ & $ 0.20 \pm 0.02$ \\
XSG1 & $0.25$ & $307 \pm 3$ &  $10.3 \pm  0.8$  &  $0.34 \pm 0.04  $ & $ 0.41 \pm 0.04$ & $ 0.19 \pm 0.03$ \\
XSG1 & $0.32$ & $302 \pm 3$ &  $10.3 \pm  0.8$  &  $0.22 \pm 0.03  $ & $ 0.35 \pm 0.03$ & $ 0.21 \pm 0.03$ \\
XSG1 & $0.43$ & $291 \pm 3$ &  $10.3 \pm  0.8$  &  $0.20 \pm 0.03  $ & $ 0.35 \pm 0.03$ & $ 0.18 \pm 0.03$ \\
XSG1 & $0.78$ & $276 \pm 3$ &  $10.3 \pm  0.8$  &  $0.08 \pm 0.03  $ & $ 0.19 \pm 0.03$ & $ 0.10 \pm 0.02$ \\
XSG2 & $0$    & $308 \pm 3$ &     $7 \pm 0.5 $  &  $0.35 \pm 0.02  $ & $ 0.27 \pm 0.03$ & $ 0.08 \pm 0.03$ \\
\hline
  \end{tabular}
\label{tab:abundances}
\end{table*}

The wide wavelength baseline provided by X-Shooter allows us to
simultaneously study all four Na-sensitive features with the same
instrument, namely the three doublets at 5890 and 5896\,\AA\ (\nai ),
at 8183 and 8195\,\AA\ (\naii ), and at 11400\,\AA (\naiii ), as well
as the line at 22100\,\AA\ (\naiv ). 
For XSG1, the measurements are
made in all six radial bins, except for \naiv, for which we omit the
measurement in the outermost bin due to the lower S/N in the K-band.
Regarding our second target, XSG2 -- for which we only use the central
spectrum -- we measure all four Na-sensitive features.  Line strengths
are computed according to the band definitions given in
Tab.~\ref{tab:defindices}. The indices \nai\ and
\naii\  are defined in \citet{Trager98}\footnote{Notice that,
  although we rely on the index definition of \citet{Trager98}, the
  indices are not measured in the Lick/IDS system, but at the effective
  resolution (including both velocity dispersion and instrumental
  resolution).}  and LB13, respectively, whereas \naiii\ follows
the definition of CvD12, although shifting the blue pseudocontinuum 
bluewards by 3\,\AA, in order to avoid a residual from data reduction, mostly
visible in the outermost spectra of XSG1, at $\lambda\!\sim\!11366$\,\AA. 
\naiv\ is based on the index defined by CvD12, but decreasing
(increasing) the lower (upper) limit of the blue (red) pseudocontinuum
by 17(12)\,\AA. This definition of \naiv\ -- featuring a wider region for the
sidebands -- reduces the statistical uncertainty on the derivation of
the pseudocontinuum, yet avoiding contamination from sky residuals,
without a significant change in the sensitivity to relevant
parameters, such as line broadening, IMF, metallicity, and abundance
ratios. This issue was verified with the aid of extended--MILES and
CvD12 stellar population models. Notice that at the redshift of our targets,
all Na features (and in particular \naii ) are not affected significantly by telluric absorption in the 
earth atmosphere. Moreover, the exquisite resolution provided by X-Shooter
allows us to achieve a sub-percent level accuracy in the removal of sky emission
and telluric lines, as detailed in LB16.

All Na-sensitive indices as well as \mgfep\ -- which are fit
simultaneously (see Sec.~\ref{sec:fitting}) -- are
corrected\footnote{Notice that we prefer this approach, instead of
  smoothing all observed spectra to a common velocity dispersion
  (typically the maximum value measured in the sample), as it extracts
  the maximum amount of information from the data (see, e.g., LB13),
  avoiding any contamination of the relevant features by neighbouring
  sky residuals.} to a common velocity dispersion $\sigma$=300\,\kms.

For each spectrum, the correction is done as follows. We compute the
difference between the line strengths at the velocity dispersion of
the data, and at the reference $\sigma$=300\,\kms, by use of the
extended--MILES models. These differences are estimated for models
with an (old) age of 11\,Gyr (i.e. typical of XSG1, see LB16), at
solar (\zh$=0$) and super-solar (\zh$=0.22$) metallicity, adopting two
choices of bimodal IMF, namely a Kroupa-like distribution
(\gammab=1.3) and a bottom-heavy IMF (\gammab=3). The median values of
these differences are then subtracted off from the observed indices
(both \mgfep and the Na-sensitive features). Notice that for all
Na-sensitive indices, the correction is largely model-independent,
differing by less than 0.015\,\AA\ among different models, i.e.
significantly below the statistical error on line strengths. This is
due to the fact that our data span a relatively narrow range in
velocity dispersion, from $\sim$270 to $\sim$330\,\kms\ (see
Tab.~\ref{tab:abundances}). The corrected values of \mgfep\ and Na
indices are reported in Tab.~\ref{tab:indices}. All the corrected
line strengths are compared to predictions of models smoothed to the
same velocity dispersion of 300\,\kms.

\begin{table*}
\centering
\small
 \caption{Index definitions of the four Na-sensitive
   features. Wavelengths are quoted in the air system.}
  \begin{tabular}{c|c|c|c|c|c}
   \hline
 Index  & Units & Blue Pseudo-continuum & Central feature & Red Pseudo-continuum & Reference \\
        &       & [\AA] & [\AA] & [\AA] &      \\
   (1)  &   (2) &   (3) &   (4) &   (5) & (6)  \\
   \hline
\nai    & \AA & $5860.625$--$5875.625$ & $5876.875$--$5909.375$ & $5922.125$--$5948.125$      & Trager+98   \\
\naii   & \AA & $8143.000$--$8153.000$ & $8180.000$--$8200.000$ & $8233.000$--$8244.000$      &  $\rm LB13$ \\
\naiii  & \AA & $11353.882$--$11363.879$& $11368.879$--$11411.867$ & $11413.866$--$11423.864$ & CvD12+this work   \\
\naiv   & \AA & $22012.000$--$22039.000$ & $22041.000$--$22099.000$ & $22100.000$--$22156.000$& CvD12+this work   \\
  \hline
  \end{tabular}
\label{tab:defindices}
\end{table*}

\begin{table*}
\centering
\small
 \caption{Measured line strengths for the total metallicity indicator
   \mgfep , and for the four Na-sensitive features analyzed in the
   present work. All line strengths have been corrected to a velocity
   dispersion of 300\,\kms\ (see Sec.~\ref{sec:indices}). Error bars
   are quoted at the 1\,$\sigma$  level.}
  \begin{tabular}{c|c|c|c|c|c|c}
  \hline
 Galaxy & $\rm R/R_e$ & \mgfep & \nai & \naii & \naiii & \naiv \\
        &             &  [\AA] & [\AA]& [\AA] & [\AA]  & [\AA] \\
  \hline
XSG1 & $0$    & $ 3.59 \pm 0.02$ & $ 5.78 \pm 0.03$ & $ 1.08 \pm 0.01$ & $ 1.40 \pm 0.05$ & $ 2.25 \pm 0.10$ \\ 
XSG1 & $0.19$ & $ 3.33 \pm 0.05$ & $ 5.24 \pm 0.07$ & $ 0.96 \pm 0.04$ & $ 1.27 \pm 0.08$ & $ 1.81 \pm 0.14$ \\ 
XSG1 & $0.25$ & $ 3.38 \pm 0.06$ & $ 5.25 \pm 0.08$ & $ 0.93 \pm 0.04$ & $ 1.27 \pm 0.10$ & $ 2.05 \pm 0.19$ \\ 
XSG1 & $0.32$ & $ 3.25 \pm 0.06$ & $ 5.12 \pm 0.08$ & $ 0.88 \pm 0.04$ & $ 0.98 \pm 0.11$ & $ 1.92 \pm 0.24$ \\ 
XSG1 & $0.43$ & $ 3.24 \pm 0.07$ & $ 4.62 \pm 0.09$ & $ 0.85 \pm 0.04$ & $ 0.97 \pm 0.15$ & $ 1.89 \pm 0.31$ \\ 
XSG1 & $0.78$ & $ 3.09 \pm 0.07$ & $ 4.25 \pm 0.10$ & $ 0.76 \pm 0.11$ & $ 0.93 \pm 0.22$ & \\
XSG2 & $0$    & $ 3.49 \pm 0.04$ & $ 5.25 \pm 0.05$ & $ 0.92 \pm 0.02$ & $ 1.06 \pm 0.10$ & $ 2.11 \pm 0.15$ \\ 
  \hline
  \end{tabular}
\label{tab:indices}
\end{table*}

\section{Abundance ratios}
\label{sec:sppars}
Since the (optical) Na features are expected to be sensitive to
\afe\ (see below), we estimate \afe\ for all spectra, relying on the
scaled-solar \afep\ proxy (see LB13 and V15). This is defined as the
difference between the total metallicity, at fixed age,
derived from either \mgb\ or \fet. 
The proxy is converted\footnote{Since the proxy is derived 
from Mg- and Fe-sensitive absorption lines, in practice
  it measures [Mg/Fe], and the conversion to \afe\ assumes a given
  ratio among different alpha elements (such as, e.g., in TMJ11, or
  V15). This assumption does
  not affect our analysis, as we treat the sensitivity of Na-dependent lines to
  \afe\ in an empirical manner (i.e. the uncertainty on the
  conversion from \mgfe\ to \afe\ is absorbed by the
  \alj\ coefficients, see Sec.~\ref{sec:fitting}).} into \afe\ with the aid of the \citet[][hereafter
  TMJ11]{TMJ11} stellar population models\footnote{Notice that, for
  the conversion of the proxy into \afe, we use TMJ11, rather than
  $\alpha$--MILES models, in order to allow for a direct comparison with
  our previous works (e.g. LB13; \citealt{NMN:15b}). However, this
  choice does not affect our conclusions, as there is a tight
  correlation between \afe\ estimates based on $\alpha$--MILES and
  TMJ11 models, as shown in figure~30 of V15.}, resulting into an
accuracy (RMS) of 0.025\,dex in \afe .  The values of \afe\ for
XSG1 and XSG2 are reported in Tab.~\ref{tab:abundances}. Notice that 
the derivation of \afe\ for XSG1 is the same as described in LB16
(see their figure~7), where we also compared different methods to
estimate \afe, finding consistent results.  XSG1 turns out to have an
almost constant \afe\ profile, with an high value of
\afe$\sim$0.4\,dex
in the center, decreasing to $\sim$0.2\,dex only in the
outermost radial bin. We find for XSG2  \afe$\sim$0.275$\pm$0.03. 
In addition to \afe, we also estimate \cfe\ (see LB16), 
as this is relevant to our discussion of abundance effects on
the NIR Na-sensitive lines (see below).  Due to the strong sensitivity of
\cfs\ to carbon abundance \citep[e.g.][]{TripiccoBell:1995}, we
estimate \cfe\ with this feature, fitting simultaneously \hbo ,
\mgfep , \mgb , \fet , and \cfs\  with the 
extended-MILES models (see Sec.~\ref{sec:models}) adopting a
Kroupa-like IMF. The fit is done by minimizing the RMS scatter between
the model and the observed line strengths with respect to \cfe\ and \afe\ (see
equation~1 of LB16), where the sensitivity of line strengths to
\afe\ (\cfe ) is computed with \amiles\ (CvD12) models, for a
13.5\,Gyr old population at solar metallicity.  As shown in LB16, the
use of models with a bottom-heavy IMF only affects the estimate of
\cfe, below 0.05\,dex. Moreover, the estimate of \afe\ from this
approach is fully consistent with that of the scaled-solar proxy. For
XSG1, the \cfe\ radial profile has a similar trend with respect to \afe , being
almost constant as a function of radius, with \cfe$\sim$0.2\,dex in
all radial bins except for the innermost and outermost bins, where 
\cfe$\sim$0.1\,dex (see figure~7 of LB16).  The same approach gives
\cfe$\sim$0.08$\pm$0.03\,dex for the central spectrum of XSG2.  Hence,
XSG2 turns out to have significantly lower abundance ratios (both
\afe\ and \cfe ) than XSG1, despite both galaxies having similar velocity
dispersion. Moreover, the central spectra of XSG1 and XSG2 have fully
consistent \afe$/$\cfe\ ratios ($\sim$3.2 and $\sim$3.4,
respectively). Notice that in the present work, the \cfe\ estimates
are used for discussion purposes only, and they do not enter the
fitting procedure of Na-sensitive line strengths.

\section{Fitting Na line strengths}
\label{sec:fitting}

\subsection{The method}
\label{sec:method}
We fit the line strengths for all the available spectra simultaneously, by
minimizing the following equation:
\begin{eqnarray}
\small
\chi^2 & = & \rm \sum_{i,j} (  Na_{j,i}^{obs} - Na_{j,i}^{mod}  
         -   \alpha_{Na_j} \cdot Na_{j,i}^{mod} \cdot [\alpha/Fe]_i  )^2 \cdot {\sigma_{j,i}}^{-2} + \nonumber \\
        & + &  \rm \sum_{i} ( [MgFe]'_{obs,i} - [MgFe]'_{mod,i} )^2 \cdot {\sigma_{[MgFe]'_i}}^{-2} 
\label{eq:chi}
 \end{eqnarray}
where the index i runs over the available spectra, the index j refers
to a given Na feature (with $\rm Na_j=\{$\nai, \naii, \naiii,
\naiv$\}$), $\rm Na_{j,i}^{obs}$ and $\rm Na_{j,i}^{mod}$ are the
observed and model Na-sensitive line strengths, $\rm \sigma_{j,i}$ are
the uncertainties on $\rm Na_{j,i}^{obs}$,
\afe$_i$ is the $\alpha$-element abundance ratio for the i-th spectrum, 
and \alj\ the {\it normalized response} of the j-th Na-sensitive index
to \afe, namely
\begin{eqnarray}
  \rm \alpha_{Na_j} & = & \rm \frac{\delta(Na_j)/Na_j}{\delta[\alpha /Fe]}. \label{eq:respal} 
  \end{eqnarray}
The last term in Eq.~\ref{eq:chi}, refers to the total metallicity
indicator \mgfep,  where $\rm [MgFe]'_{obs,i}$ and  $\rm [MgFe]'_{mod,i}$ 
are the observed and model line-strengths, respectively, and 
$\rm \sigma_{[MgFe]'_i}$ corresponds to the uncertainties on 
$\rm [MgFe]'_{obs,i}$. 
{ Notice that the $\rm Na_{j,i}^{obs}$ are measured for the central spectrum of XSG2 as well as 
for all six radial bins of XSG1, except \naiv\ in the outermost bin.
Hence, the $\chi^2$ in Eq.~\ref{eq:chi} includes a total of $34$ terms, i.e. 
$6\times 4 -1$ (XSG1) plus 4 (XSG2) constraints from Na features, along with $7$ additional 
constraints from $\rm [MgFe]'_{obs,i}$. The free-fitting parameters in Eq.~\ref{eq:chi} are 
the \{\alj\}'s, and -- for each spectrum -- its IMF slope, the \nafe$_i$, and \zh$_i$. This 
amounts to $4+7+7+7=25$ free-fitting parameters, plus two constrained best-fit parameters 
for age (see Sec.~\ref{sec:constraints} for details). 
Notice also that while 
for each fit the values of \afe$_i$ are kept fixed in Eq.~\ref{eq:chi}, the uncertainties on \afe$_i$ 
are correctly propagated to those on best-fitting parameters (see end of Sec.~\ref{sec:constraints}).
Moreover, at each step of the $\chi^2$ minimization procedure, the values of $\rm \sigma_{j,i}$ in 
Eq.~\ref{eq:chi} are updated~\footnote{ In practice, this is done by adding in quadrature to 
the uncertainty on $\rm Na_{j,i}^{obs}$, the term 
$\rm \alpha_{Na_j} \cdot Na_{j,i}^{mod} \cdot \sigma_{[\alpha/Fe]_i}$, where $\rm \sigma_{[\alpha/Fe]_i}$  
is the measurement error on \afe$_i$ (see Tab.~\ref{tab:abundances}).
}, in order to include also the error budget due 
to the uncertainties on
\afe$_i$. 
}
All model line strengths in Eq.~\ref{eq:chi} 
are derived from the Na-MILES SSP models, and depend on age, metallicity, 
IMF, and  \nafe. Notice that  stellar age is treated as a 
free-fitting parameter, assuming suitable priors/constraints, as detailed in
Sec.~\ref{sec:assumptions}. We prefer not to include any age
indicator (i.e. Balmer lines) in Eq.~\ref{eq:chi}, in order to avoid
issues related to their sensitivity to the IMF\footnote{For instance, as
  discussed in V15, the dependence of \hbo\ on the IMF is model-dependent:
  while the V15 models give a decrease with IMF slope, the CvD12 models
  show a slight increase.}, as well as issues
related to emission correction, dependence of Balmer lines on
abundance ratios, and uncertainties on the absolute zero-point of the
age from different indicators (see also Sec.~\ref{sec:constraints}).
Nevertheless, we confirm that the inclusion of \hbo\ in the analysis
does not change significantly the results.

\subsection{Assumptions}
\label{sec:assumptions}
The terms $\rm \alpha_{Na_j} \cdot Na_{j,i}^{mod} \cdot [\alpha/Fe]_i$
in Eq.~\ref{eq:chi} describe the response of the j-th Na-sensitive
feature to \afe\ for the i-th spectrum, and account for the fact that
Na-MILES models have not been computed for a varying \afe.  More
importantly, we aim to constrain the effect of \afe\ on an empirical
basis, relying as much as possible on the data. Hence, in our
approach, we treat the terms \alj\ as free fitting parameters.
To this effect, we assume that the coefficients \alj are constant,
i.e. independent\footnote{This assumption is adopted within the range of
  age and metallicity of our targets, i.e. old and metal-rich
  populations ($\gtrsim$8\,Gyr; \zh$\gtrsim$0.} of age, metallicity, and IMF, as
well as any other stellar population properties (e.g. abundance ratios).
This assumption is supported for \nai\ by the
predictions of $\alpha$--MILES SSP models (see V15), as shown in
Tab.~\ref{tab:amilesresp}, where we report the values of $\rm
\alpha_{NaD} = \rm \frac{\delta(Na_D)/Na_D}{\delta[\alpha /Fe]}$ for
models with different age, metallicity, and (bimodal) IMF slope. The
derived $\rm \alpha_{NaD}$ is roughly constant, with a median value of 
$-0.83$\,dex$^{-1}$.  Regarding the other Na-sensitive features,
Tab.~\ref{tab:cvdresp} reports the values of \alj\ for a CvD12 
13.5\,Gyr SSP with solar metallicity, and a Chabrier IMF. Hereafter,
we refer to these values as \alj$\rm _{CvD}$.  Notice that
for the NaD index, $\rm \alpha_{NaD,CvD} \sim\!-0.6 \, dex^{-1}$, which
is significantly lower than the value from the $\alpha$-MILES models.  This
discrepancy is at least partly due to the fact that the CvD12 models are
computed at fixed [Fe/H] metallicity (see V15 for details).
Increasing \afe\ in the CvD12 models results in an increase in \zh. 
Since NaD increases with \zh, the decrease of NaD with \afe\ 
is expected to be more pronounced at fixed \zh\ (i.e. $\alpha$-MILES) 
than at fixed [Fe/H] (i.e. CvD12). 
Notice that according to the CvD12 models, one expects a very weak
sensitivity of the NIR \naiii\ and \naiv\ indices to \afe, making our
assumption of constant \alj\ for \naiii\ and \naiv\  more accurate.
CvD12 predict a higher sensitivity to \afe\ of \naii\ with respect to 
\nai. Although, we cannot test our assumption of a 
constant \alj\ for \naii, we point out that under
this assumption we are able to fit all Na features simultaneously for
all available spectra. Moreover, the best-fit value of 
$\rm\alpha_{NaD}$ turns out to be remarkably consistent with that expected
from the $\alpha$-MILES models, as discussed in Sec.~\ref{sec:results}.


\subsection{Fitting parameters and additional constraints}
\label{sec:constraints}
{ Our analysis relies on a total of $34$ observed data points 
(see Sec.~\ref{sec:method}), and 
$4+7+7+7=25$ free-fitting parameters (i.e. {  \{\alj\} }, the IMF slopes, \{\nafe$_i$\}, and
\{\zh$_i$\}; see Eq.~\ref{eq:chi}) plus two constrained best-fit 
parameters related to age (see below).
} 
However, as described in Sec.~\ref{sec:cases}, we also
consider fits where the number of free parameters is significantly reduced, either
by keeping fixed some of the \{\alj\} or the IMF slopes fixed\footnote{
Moreover, as mentioned in Sec.~\ref{sec:naetgs}, we also get consistent results to those 
presented in this paper when 
including \nai\ and \naii\ line-strengths for the SDSS stacked with 300~\kms\ from LB13, 
in our fitting approach.
}. In
these cases, we still find consistent results with respect to the
general case. The minimization of Eq.~\ref{eq:chi} is performed by
computing $\chi^2$ over a grid of models with varying age,
metallicity, IMF, and \nafe , deriving the correction factors\{\alj\}
that minimize $\chi^2$, for each position in the grid, based on a
Levenberg-Marquardt algorithm. Notice that while metallicity and IMF
are free-fitting parameters for each spectra, the age is allowed to
vary between 9 and 12\,Gyr for XSG1, i.e. following the constraints
obtained with different methods (see section~4.3 of LB16). 
In the case of XSG2, we assume a gaussian constraint on the age
of 8$\pm$1.6\,Gyr, inferred by comparing different methods (e.g. \hbo\ and full
spectral fitting), as in XSG1 (see LB16). We impose as a
further constraint, that stellar age is the same for all spectra of XSG1,
since this galaxy exhibits no significant age gradient (see LB16).
These assumptions introduce two
further constraints, i.e. the age of (all spectra) for XSG1 and the
age of the central spectrum of XSG2.  For completeness, we confirm
that leaving the age parameter completely unconstrained for all
spectra does not introduce significant changes.  Notice that there
is a substantial uncertainty in the absolute determination of the age
among different indicators (see, e.g., LB16). Thus, as mentioned in
Sec.~\ref{sec:fitting}, we do not explicitly include any
age-sensitive feature, such as \hb, in Eq.~\ref{eq:chi}, 
adopting instead the age constraints that account for differences among
different methods.
Regarding the \{$\rm [\alpha/Fe]_i$\} terms in Eq.~\ref{eq:chi}, we keep
them fixed to values computed from the optical lines (see
Sec.~\ref{sec:indices}, and Tab.~\ref{tab:abundances}).  Uncertainties
on best-fit parameters in Eq.~\ref{eq:chi} are estimated from
$N=1000$ bootstrap iterations, where the fitting is repeated after
shifting the observed line strengths and the values of \{$\rm [\alpha/Fe]_i$\}
according to their uncertainties.


\begin{table}
\centering
\small
 \caption{Normalized response of \nai\ to \afe\ from $\alpha$--MILES stellar population models,
 with varying IMF (col.~1), age (col.~2), and metallicity (col.~3). 
 Absolute (i.e. $\rm \delta NaD/ \delta[\alpha/Fe]$) and relative (i.e. $\rm \delta NaD/ NaD / \delta[\alpha/Fe]$) 
 responses to \afe\  are listed in cols.~4 and~5, respectively.}
  \begin{tabular}{c|c|c|c|c}
   \hline
  IMF &   Age   &   \zh\ & $\rm \frac{\delta NaD}{\delta[\alpha/Fe]}$ &   $\rm \frac{\delta NaD/ NaD}{\delta[\alpha/Fe]}$  \\
      & (Gyr) & dex  & $\rm \AA/dex$ & $\rm dex^{-1}$ \\
  (1)   &      (2)       &       (3)    & (4) & (5)  \\
   \hline
$ \rm \Gamma_b=1.30 $ & $    8 $ & $ \rm 0.06 $ & $   -2.44$ & $   -0.83 $ \\ 
$ \rm \Gamma_b=1.30 $ & $   14 $ & $ \rm 0.06 $ & $   -2.69$ & $   -0.81 $ \\ 
$ \rm \Gamma_b=3.30 $ & $    8 $ & $ \rm 0.06 $ & $   -3.30$ & $   -0.87 $ \\ 
$ \rm \Gamma_b=3.30 $ & $   14 $ & $ \rm 0.06 $ & $   -3.57$ & $   -0.85 $ \\ 
$ \rm \Gamma  =2.30 $ & $    8 $ & $ \rm 0.06 $ & $   -3.04$ & $   -0.86 $ \\ 
$ \rm \Gamma  =2.30 $ & $   14 $ & $ \rm 0.06 $ & $   -3.39$ & $   -0.85 $ \\ 
$ \rm \Gamma_b=1.30 $ & $    8 $ & $ \rm 0.26 $ & $   -2.91$ & $   -0.81 $ \\ 
$ \rm \Gamma_b=1.30 $ & $   14 $ & $ \rm 0.26 $ & $   -3.40$ & $   -0.83 $ \\ 
$ \rm \Gamma_b=3.30 $ & $    8 $ & $ \rm 0.26 $ & $   -3.72$ & $   -0.84 $ \\ 
$ \rm \Gamma_b=3.30 $ & $   14 $ & $ \rm 0.26 $ & $   -4.16$ & $   -0.84 $ \\ 
$ \rm \Gamma  =2.30 $ & $    8 $ & $ \rm 0.26 $ & $   -3.40$ & $   -0.83 $ \\ 
$ \rm \Gamma  =2.30 $ & $   14 $ & $ \rm 0.26 $ & $   -3.94$ & $   -0.84 $ \\ 
  \hline
  \end{tabular}
\label{tab:amilesresp}
\end{table}


\begin{table}
\centering
\small
 \caption{Normalized response of Na features to \afe\ (see Eq.~\ref{eq:respal}) from CvD12 SSP models,
 with an age of 13.5~Gyr, solar metallicity, and a Chabrier IMF.}
  \begin{tabular}{c|c}
   \hline
 Index  &      \alj$\rm _{CvD}$  \\
        & $\rm dex^{-1}$ \\
  (1)   &      (2)       \\
   \hline
 \nai     & -0.59  \\
 \naii    & -0.83  \\ 
 \naiii   & -0.07  \\ 
 \naiv    & -0.20  \\ 
  \hline
  \end{tabular}
\label{tab:cvdresp}
\end{table}

\subsection{Fitting cases}
\label{sec:cases}

We use Eq.~\ref{eq:chi} to constrain the relevant stellar population
parameters, i.e.  the relative response of the Na-sensitive features
to \afe, as well as the IMF slopes and \nafe\ abundance ratios for
XSG1 and XSG2.  In order to test the robustness of our results, the
fitting is done for different cases, as summarized in
Tab.~\ref{tab:methods}. In all cases, we fit data for both galaxies
simultaneously, while we consider different options. Notice that in
all figures, throughout the present work, we apply the same colour
coding for results derived from the same fitting scheme, as follows
(see also Tab.~\ref{tab:methods}):

\begin{description}
 \item[ (A; green) -- ] We set the \afe\ sensitivity of the NIR Na
   features (\naiii\ and \naiv ) to zero, based on the prediction from
   the CvD12 models (see Tab.~\ref{tab:cvdresp}, and
   Sec.~\ref{sec:fitting}). Additionally, we impose gaussian
   constraints on the IMF slope of XSG1 for each radial bin.  This is
   done by adding the term $\rm \sum_i [\Gamma_{b,i} -
     \Gamma_{b,i,LB16}]^2/\sigma^2_{\Gamma_{b,i}}$ in Eq.~\ref{eq:chi},
   where $\rm \Gamma_{b,i,LB16}$ and $\rm \sigma_{\Gamma_{b,i}}$ are
   the IMF slope value derived for the i-th radial bin of XSG1 from
   LB16, and its 1\,$\sigma$ uncertainty, while $\rm \Gamma_{b,i}$ is the
   IMF slope of the SSP model for which model line strengths ($\rm
   Na_{j,i}^{mod}$ and $\rm [MgFe]'_{mod,i}$) are computed at each
   step of the minimization procedure. Notice that the IMF slope of
   XSG2 is treated as a free fitting parameter.
 \item[ (B; red) -- ] Same as case (A), but without any
   constraint on the IMF for XSG1.
 \item[ (C; black) -- ] Same as case (A), but without including
   \naiv\ in the fit. This is motivated by the fact that the
   \naiv\ line might be sensitive to other abundance
   ratios (e.g. \cfe, see \citealt{ester:2009}; R\"ock et al., in
   preparation), besides \nafe.
 \item[ (D; blue) -- ] All correction coefficients \{\alj\} are
   treated as free-fitting parameters, but IMF
   constraints are applied for XSG1.
 \item[ (E; magenta) -- ] Same as case (D), but without any
   constraint on the IMF of XSG1.
 \end{description}

\begin{table*}
\centering \small
\caption{Different fitting assumptions used in this paper, labelled by
  a letter (column~1), and shown in the figures with a { colour and line type as given
  in columns~5 and~6, respectively}. Among different cases, we vary the set of \alj\ coefficients
  that are fitted to the data (column~2; see Eq.~\ref{eq:chi}), the
  use of IMF constraints for XSG1 from LB16 (column~2), and the set of
  fitted Na-sensitive features (column~4).  }
  \begin{tabular}{c|c|c|c|c|c}
  \hline
 label &            \alj               & IMF constraints for XSG1   & Na features & colour & line type \\
  (1)  &             (2)               &            (3)              &     (4)     &   (5)  & (6) \\
 \hline
   (A) &   $\alpha_3=0$, $\alpha_4=0$  &           yes               &     all     &  green   & solid \\
   (B) &   $\alpha_3=0$, $\alpha_4=0$  &           no                &     all     &  red     & dashed \\
   (C) &   $\alpha_3=0$, $\alpha_4=0$  &           yes               &  no \naiv\  &  black   & dot-dashed \\
   (D) &           all free            &           yes               &     all     &  blue    & dotted \\
   (E) &           all free            &           no                &     all     &  magenta & dash with three dots\\
  \hline
  \end{tabular}
\label{tab:methods}
\end{table*}

\begin{figure*}
\begin{center}
\leavevmode
\includegraphics[width=17cm]{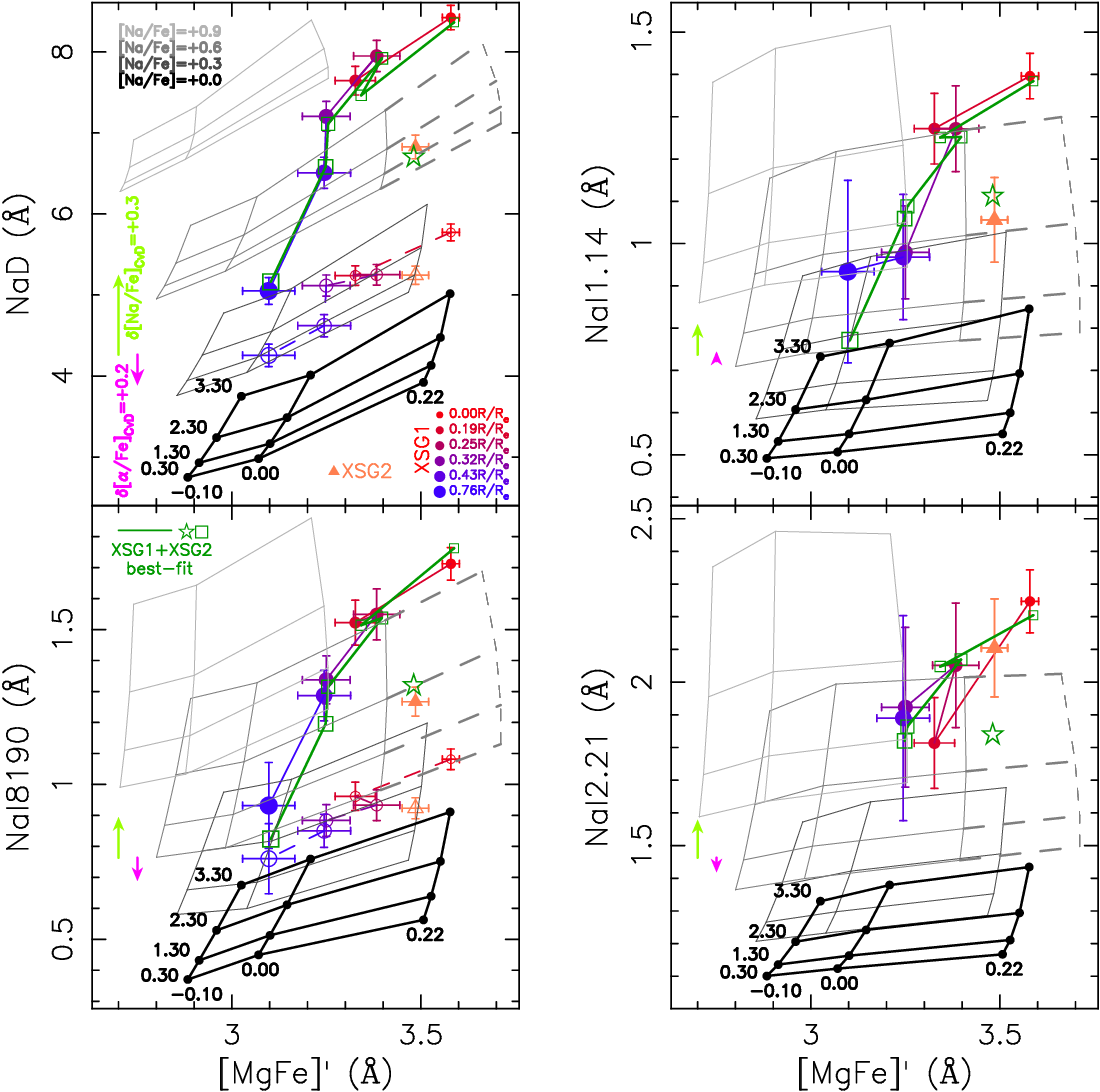}
\end{center}
\caption{Line strengths of Na-sensitive features as a function of the
  total metallicity indicator \mgfep . Data-points for XSG1 are
  plotted as circles, and colour-coded from red through blue, as a
  function of galacto-centric distance (see labels in the lower--right
  corner of top--left panel), while data-points for XSG2 are shown as
  orange triangles. { Symbol sizes increase with galacto-centric distance}. 
  In the panels for \nai\ and \naii\, filled
  circles (connected by a solid line) and triangles plot line
  strengths corrected to \afe$=0$, for XSG1 and XSG2, respectively,
  while empty circles (connected by a dashed line) and triangles plot
  observed line strengths without any correction applied.  The
  corrections are negligible for \naiii\ and \naiv, as these indices
  are not sensitive to \afe .  The black grids, and black dots, show
  predictions of extended--MILES SSP models, with varying metallicity
  (\zh$\rm =-0.1, 0, +0.22$), bimodal IMF slope (\gammab$=+0.3, 1.30,
  2.30, 3.30$), and \nafe$=0$.  Grey grids show predictions of
  Na-MILES models with \nafe$=0.3$, 0.6, and 0.9, respectively (see
  labels in the top--left corner). For reference, \gammab$\sim 1.3$
  corresponds approximately to the case of a Kroupa IMF.  All grids
  are computed for an age of $11.2$~Gyr. For the grids with
  \nafe$=+0.6$, we also show the linear extrapolation of the models,
  up to \zh$\rm =+0.4$, performed to match \mgfep\ for the innermost
  data-points of XSG1 and XSG2 (see the text).  In each panel,
  light-green and magenta arrows show the effect of increasing
  \afe\ (\nafe) by $+0.2$~dex ($+0.3$~dex) on Na line strengths,
  according to CvD12 model predictions, computed for an age of
  13.5~Gyr, solar metallicity, and a Chabrier IMF.  Notice that line
  strengths corrected to \afe=0 (filled circles and triangles) can be
  directly compared to Na-MILES model predictions (i.e. black and grey
  grids).  Green squares, connected by a solid line, and green stars
  are best-fit results for XSG1 and XSG2, respectively, for our
  reference fitting approach (case A, see Sec.~\ref{sec:cases}) where
  we constrain the IMF profile of XSG1 to be the same as that derived
  in our previous work (LB16). Notice that we are able to match all
  Na-sensitive line strengths in both galaxies.
}
\label{fig:na_fits}
\end{figure*}

\begin{figure}
\begin{center}
\leavevmode
\includegraphics[width=8.5cm]{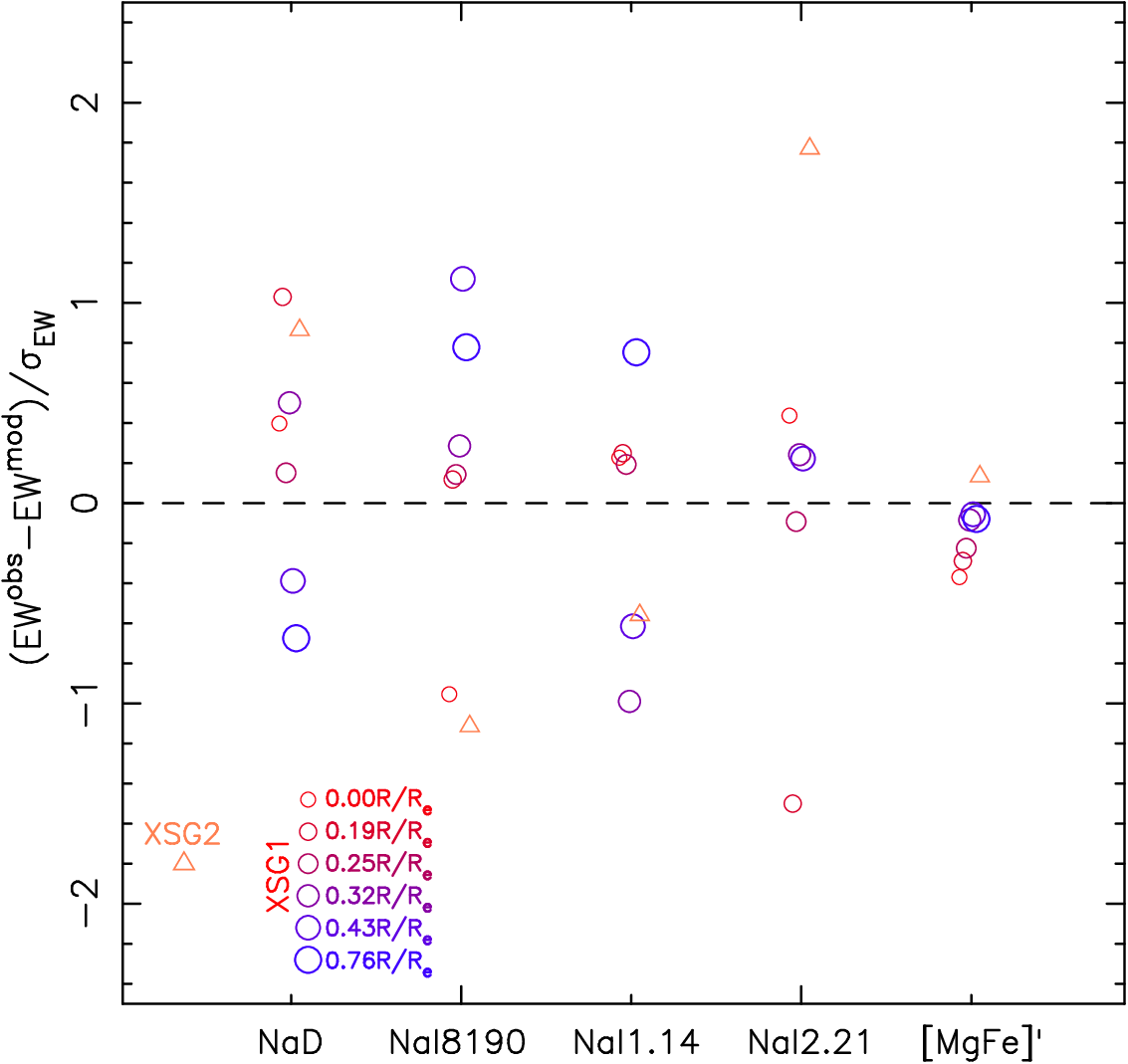}
\end{center}
\caption{ Differences between observed and best-fit line-strengths, normalized to their 
uncertainties, for all features and spectra included in our fitting procedures,
using the fiducial approach (case A, see Tab.~\ref{tab:methods}), where we
impose IMF constraints on XSG1 from LB16. The horizontal dashed line marks the value of 
zero. Symbol types and colours for XSG1 and XSG2 are the same as in Fig.~\ref{fig:na_fits}.
Notice that all features are fitted remarkably  well, with typical differences 
$\simlt1$\,$\sigma$. 
}
\label{fig:na_disc}
\end{figure}

\section{Results}
\label{sec:results} 

Fig.~\ref{fig:na_fits} shows the observed line strengths of
Na-sensitive features in XSG1 (circles) and XSG2 (triangles), along
with the best fit results from our SSP models that include a varying
metallicity, (bimodal) IMF, and [Na/Fe].  All four Na-sensitive indices
are shown as a function of the total metallicity indicator \mgfep.
In a recent paper (LB16) we could rule out a single power law
(unimodal) functional form for the IMF in XSG1, on the basis of TiO and FeH
gravity-sensitive indices. Hence, in the present work we only consider
a bimodal distribution.  The black grids and solid dots in the Figure show the
response of the indices to \zh\ and IMF slope -- derived from the extended--MILES models,
i.e. at \nafe$\sim$0 -- while the grey grids correspond to (theoretical)
models with \nafe=$+$0.3, $+$0.6, and $+$0.9\,dex, respectively.  Notice that
models with \nafe=$+$1.2\,dex have also been computed (see
Sec.~\ref{sec:models}), and used in the fitting procedure\footnote{In
  practice, we allow \nafe\ to vary up to $+$1.3\,dex, performing a linear
  (mild) extrapolation of model predictions above \nafe=$+$1.2\,dex.}.
However, we do not show these high values in Fig.~\ref{fig:na_fits},
to avoid overly increasing the range of index values along the
vertical axes. The observed line strengths of \nai\ and \naii\
(empty circles and triangles) are corrected to \afe=0 (filled
circles and triangles), in order to allow for a direct comparison with the
black and grey grids~\footnote{
The corrections are performed using best-fitting values of 
$\rm\alpha_{NaD}$ and $\rm\alpha_{NaI8190}$ from method A (see Tab.~\ref{tab:methods}).
However, they are approximately independent of the fitting methodology. as all methods
give consistent estimates of $\rm\alpha_{NaD}$ and $\rm\alpha_{NaI8190}$ (see Sec.~\ref{sec:aferesponses}).
}. Notice that the correction is negligible for
\naiii\ and \naiv\ (see below).  All model grids in the Figure refer
to the same (old) age of 11.2\,Gyr, with line strengths given at a
resolution of $\sigma$=300\,\kms\ (see Sec.~\ref{sec:indices}).
\begin{figure}
\begin{center}
\leavevmode
\includegraphics[width=8.2cm]{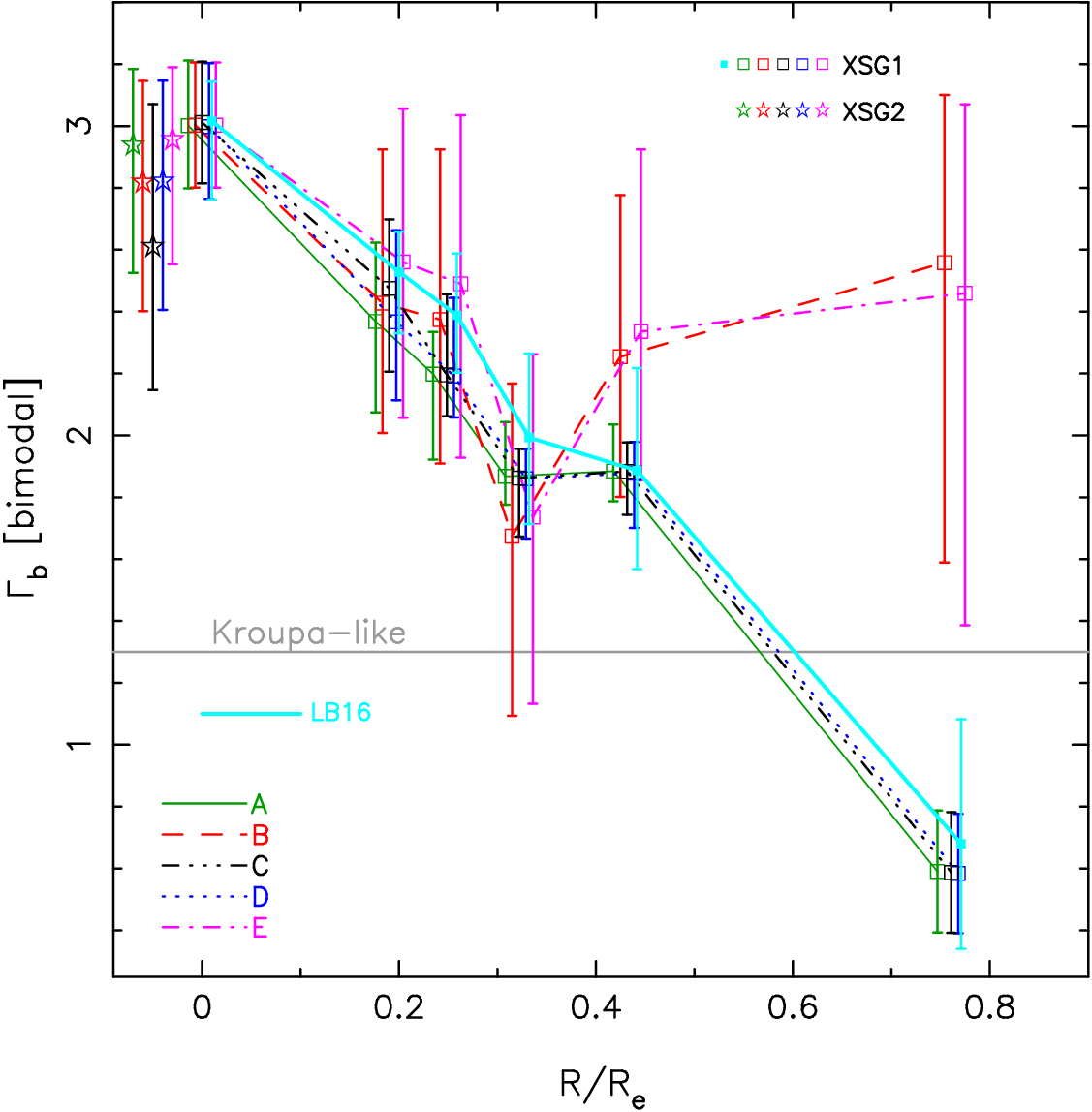}
\end{center}
\caption{IMF slopes for different radial bins of XSG1 (squares) and
  the central spectrum of XSG2 (stars).  The { cyan, thick solid}, profile corresponds
  to the radial slope of the IMF in XSG1 from LB16, that combined
  constraints from TiO features and the Wing-Ford band.  Different
  fitting cases (see Sec.~\ref{sec:cases}) are { colour- and type-coded} (see
  lower--left corner in the Figure), labelled as in
  Tab.~\ref{tab:methods}. The error bars are shown at the 1\,$\sigma$ level.
  Notice that horizontal shifts have been applied among different
  fitting cases, for ease of visualization { (the shift increasing from case A through E)}.  
  The horizontal grey line
  corresponds to a Kroupa-like IMF. Notice that in cases where no additional IMF
  constraints on XSG1 are applied (magenta and red curves), the
  best-fit IMF slopes of XSG1, based on Na-sensitive features, are consistent
  with those derived by LB16, except for the outermost radial bin where,
  as shown by the large error bars, the IMF slope is poorly
  constrained { (see the text)}.  }
\label{fig:imf_rad_na}
\end{figure}

\subsection{Behaviour of the models}
\label{sec:modelsbehaviour}
As seen in Fig.~\ref{fig:na_fits}, Na-MILES models predict that {\it
  all} four Na-sensitive lines increase significantly with IMF slope
{\it and} \nafe.  \nai\ is the most sensitive index to \nafe, but it
also increases significantly with IMF slope, as already pointed out by
LB13.  Among the ``red'' Na lines (\naii, \naiii, and \naiv ),
\naiii\ is the most sensitive one to IMF variations, followed by
\naii\ and \naiv . An important outcome is that the response of red Na
lines to \nafe\ is strongly coupled with IMF variations. For a
bottom-heavy distribution, the line strengths of red Na-sensitive
features increase more with \nafe, with respect to a Kroupa-like
distribution. This trend effectively ``stretches'' the grids in
Fig.~\ref{fig:na_fits} along the IMF (vertical) axis for higher values
of \nafe. This result is due to the fact that, for a given increase of
\nafe, the variation of Na-sensitive line strengths in cool dwarf
stars is larger than in other types of stars, making a bottom-heavy
IMF boost up the effect of \nafe\ at the level of SSP model
predictions. For \naii , a similar result was obtained by LB13 (see
their figure~15), who found that the response of \naii\ to \nafe, for
a Salpeter IMF, is larger than that for a Kroupa-like distribution,
based on an independent set of (fully theoretical) stellar population
models.  The behaviour of Na-MILES models is similar to that of CvD12
models, as both sets of models predict \nai\ to be very sensitive to
\nafe\ (see light--green arrows in Fig.~\ref{fig:na_fits},
corresponding to an increase of 0.3\,dex in \nafe\ for CvD12 models)
and \naiii\ to be the line most sensitive to IMF (see figure~12 of
CvD12). However, in contrast to CvD12, who found that the sensitivity
of \naii\ and \naiii\ to IMF is far larger than that to \nafe, we find
that the effect of \nafe\ is quite significant for {\it all}
Na lines.  Moreover, as one can see by comparing the size of
light--green arrows in Fig.~\ref{fig:na_fits} to the offset between
grids with \nafe$\sim$0 (black) and \nafe=$+$0.3, our models predict a
stronger increase with Na abundance than the CvD12 models. Despite
these differences, as shown in App.~\ref{app:rescaled}, our results
remain essentially unchanged if one matches the IMF sensitivity of the
Na line strengths of our models to those from CvD12, in which case one
derives mildly lower IMF slopes and higher \nafe\ abundances, in the
central regions.  We emphasize that while in the publicly available
version of CvD12 models, the effect of \nafe\ is only computed at a
given locus of parameter space (i.e. for a given age, metallicity, and
for a Chabrier IMF), our new set of models are the first ones
providing SSP model predictions for different \nafe\ ratios, over a
wide range of age, metallicity, and IMF slope.  Another interesting
aspect of Fig.~\ref{fig:na_fits} is that an increase of \nafe\ shifts
the model grids towards lower values of \mgfep.  In other words,
\mgfep\ decreases with \nafe . This is somewhat expected, as Na is a
major electron donor, and thus it affects electron pressure and
ionization balance in the stellar atmospheres. For instance, the line
strength of the NIR Calcium triplet lines decreases with \nafe\ (see
CvD12). Similarly, increasing \nafe\ makes the optical Fe and Mgb line
strengths to diminish, resulting in a lower \mgfep. This is relevant
as in most stellar population studies, \mgfep\ is adopted as a
total-metallicity indicator, and is assumed to be independent of
abundance ratios.  While this is approximately true for
\afe\ (\citealt{TMJ11}, V15), the present work shows that this is not
the case for \nafe.  Hence, for an Na-enhanced stellar population --
as in the massive ETGs analyzed in this work -- one would
tend to underestimate total metallicity when relying on scaled-solar
SSP models. This explains why our best-fit estimates of \zh\ for
XSG1, based on Na-enhanced models (see Tab.~\ref{tab:abundances}), are
larger, by about 0.1\,dex, than those obtained with scaled-solar
models (see figure~5 of LB16).

\subsection{Comparison of data to model grids}
\label{sec:datamodels}
Note that in Fig.~\ref{fig:na_fits} all data points remarkably lie
above the model grids of the extended--MILES models (black grids with
dots in the Figure). Therefore, we can robustly confirm that no SSP
model can match the observations with \nafe$\sim$0, regardless of the
adopted IMF, even in the outermost radial bin of XSG1, where
a solar metallicity is found (see LB16). Regarding \nai\ and \naii , whose
line strengths are anticorrelated with \afe\ (see CvD12,
\citealt{spiniello:2015}), the offset is significantly lower than in
the NIR Na-sensitive lines (\naiii\ and \naiv), for which the effect of 
\afe\ -- according to the CvD12 models -- is expected to be negligible.
Qualitatively, the behaviour of the four Na-sensitive lines implies that,
in addition to any possible IMF variation, \nafe\ has to be significantly enhanced in 
both XSG1, at all radii, and XSG2. In the optical lines, the
effect of \nafe\ is (partly) counteracted by \afe, reducing
the mismatch from the \nafe=0 grids.  This can be seen by a
comparison of the positions of the empty circles and triangles in the left
panels of Fig.~\ref{fig:na_fits} -- corresponding to the observed \nai\ and
\naii\ line strengths for XSG1 and XSG2, respectively -- to the filled
circles and triangles -- that represent the line strengths after being corrected
to \afe=0,
using the best-fit values of \alj\ (see Eq.~\ref{eq:chi}), for method
A (see Tab.~\ref{tab:methods}). Notice also that because the model
\mgfep\ decreases with \nafe\ (see above), in order to match models
with \nafe$>0$ to the (high) values of \mgfep\ for the innermost bins
of XSG1 and XSG2 , we need some extrapolation of the model grids to
the high metallicity regime, above the maximum value of \zh$=0.22$~dex
provided by MILES.  To this effect, in the fitting procedure, all
models are linearly extrapolated~\footnote{We also re-ran the fits bu extrapolating the models
up to $\zh \sim 0.6$, finding no significant differences with respect to those reported throughout the paper.} up 
to $\zh \sim 0.4$~dex.  Extrapolated
models are shown in Fig.~\ref{fig:na_fits} for models with
\nafe$=+0.6$~dex.  Because of the smooth shape of the grids in
Fig.~\ref{fig:na_fits}, the effect of extrapolation is mild, and
affects only the analysis of the innermost spectra of our
targets. Moreover, as discussed in Sec.~\ref{sec:upperlimits}, our
conclusions do not change if we neglect the dependence of \mgfep\ on
\nafe\ (i.e. if we impose all grids in Fig.~\ref{fig:na_fits} to have
the same \mgfep\ as for \nafe$=0$), in which case no extrapolation of
the models is required.

\subsection{Best-fit line strengths}
\label{sec:bestfits}
The dark-green squares and stars in Fig.~\ref{fig:na_fits} show the
best-fit line strengths for XSG1 and XSG2, respectively, using the
fiducial approach (case A, see Tab.~\ref{tab:methods}), where we
impose IMF constraints on XSG1 from LB16. Regarding \nai\ and \naii,
the best-fit line strengths are corrected to \afe=0, and should be
compared to the \afe--corrected data (filled circles and triangles;
for XSG1 and XSG2, respectively). The comparison shows a remarkable agreement 
between best-fit and observed line-strengths. 
{ 
This is shown more explicitly in Fig.~\ref{fig:na_disc}, where we plot 
the difference of observed and best-fit line-strengths,
normalized to their uncertainties, for all spectra and all 
features included in our fitting procedure.
{\it Our new Na-enhanced SSP models allow us to simultaneously match,
  at $\sim $1\,$\sigma$ level, all four Na features, in both galaxies}.
For \naiv , a mild difference (only at $1.5$\,$\sigma$) is found 
for the second radial bin of XSG1, while a more significant difference
is found for the central bin of XSG2, for which the best-fit
value is consistent with the observations only at the
1.8\,$\sigma$ level} (however, note the comparatively large
uncertainty on \naiv\ for XSG2).  Another possible explanation for { this
marginal discrepancy } is that \naiv\ is significantly affected by other
abundance ratios.  For instance, CvD12 models with super-solar [Si/Fe]
cause an increase of \naiv, consistent with the fact that Si, 
an $\alpha$ element, is a significant absorber in the region where
\naiv\ is defined \citep{S08}. Moreover, \citet{ester:2009} reported a
significant correlation of \naiv\ with the optical \cfs\ index in
ETGs, implying a possible contribution of \cfe\ to the index (R\"ock
et al., in preparation). If this were the case, since XSG2 has lower
\afe\ and \cfe\ than XSG1 (see Tab.~\ref{tab:abundances}), we would
expect a weaker \naiv\ than the best-fit solution, in contrast with
the result shown in Fig.~\ref{fig:na_fits} { and Fig.~\ref{fig:na_disc}. Furthermore, we are able
to fit remarkably well the XSG1 data, including \naiv, at 
both the innermost and outermost radii}.  We notice that all the fitting scenarios 
(see Tab.~\ref{tab:methods}) work equally well. To our knowledge,
this is the first time that all the Na-sensitive line strengths
in massive ETGs are matched simultaneously, in a consistent manner,
from the optical to the NIR.

\subsection{Best-fit IMF slopes}
\label{sec:IMFslopes}
Fig.~\ref{fig:imf_rad_na} compares the best-fit IMF slopes in XSG1 and
XSG2 for the different fitting scenarios (see \S\ref{sec:cases}).
Remarkably, when the IMF slopes in XSG1 are treated as free parameters,
without any IMF constraint (red and magenta curves), the fit
gives a bottom-heavy distribution in the innermost radial bin of this galaxy.
In the outer regions, the
values of \gammab\ are fairly consistent with the profile derived by
LB16 based on different IMF indicators (cyan curve in the
Figure){,  although for the outermost radial bin, the Na-based  IMF slope for XSG1
tends to be higher than that from LB16. 
Notice that this does {\it not} imply an inconsistency among Na features
and other IMF indicators (i.e. TiO and FeH; see LB16).
In the outermost radial bin, the uncertainty on IMF slope in XSG1 is too large to provide a significant
constraint, being consistent at $\sim 1.2$\,$\sigma$ with either a bottom-heavy or a Kroupa-like 
($\rm \Gamma_b \sim 1.3$) slope. Hence, within error bars, the Na-based IMF 
trend for XSG1 is also consistent with the finding of LB16, i.e. a variation from bottom-heavy in the center 
to Kroupa-like beyond $\sim 0.5$\re .
In fact, when
imposing additional IMF constraints on XSG1 (green, black, and blue
curves), the inferred IMF profile is very consistent with that of
LB16, {with all Na features in the innermost and outermost bins being matched 
well, at $\sim 1$\,$\sigma$, from the best-fit models (see Fig.~\ref{fig:na_disc}).}
The large uncertainty on IMF slope in the outermost radial bin
mostly reflects the large error bar on \naiii\ (see Fig.~\ref{fig:na_fits}), and the lack
of \naiv . On the other hand, the \nafe\ is better constrained (see below), thanks 
to the smaller uncertainties on \nai\ (and \naii ).
Notice also that, as discussed in~\citet{NMN:15a}, for a ``light'' IMF 
($1\simlt\Gamma_b\simlt2$) it is far more difficult to obtain a fully consistent 
IMF determination among different indicators, as the IMF effect on spectral indices becomes
subdominant wrt that of other relevant parameters (e.g. abundance ratios).}

{ For what concerns XSG2}, our analysis also points to a bottom-heavy distribution
in the central region. However, in this case, the data are also
consistent with a lighter distribution, with \gammab$\sim$2.2, or even
lower (see black star with error bar in Fig.~\ref{fig:imf_rad_na}).

\subsection{Best-fit \nafe}
\label{sec:naferatios}
\begin{figure}
\begin{center}
\leavevmode
\includegraphics[width=8.2cm]{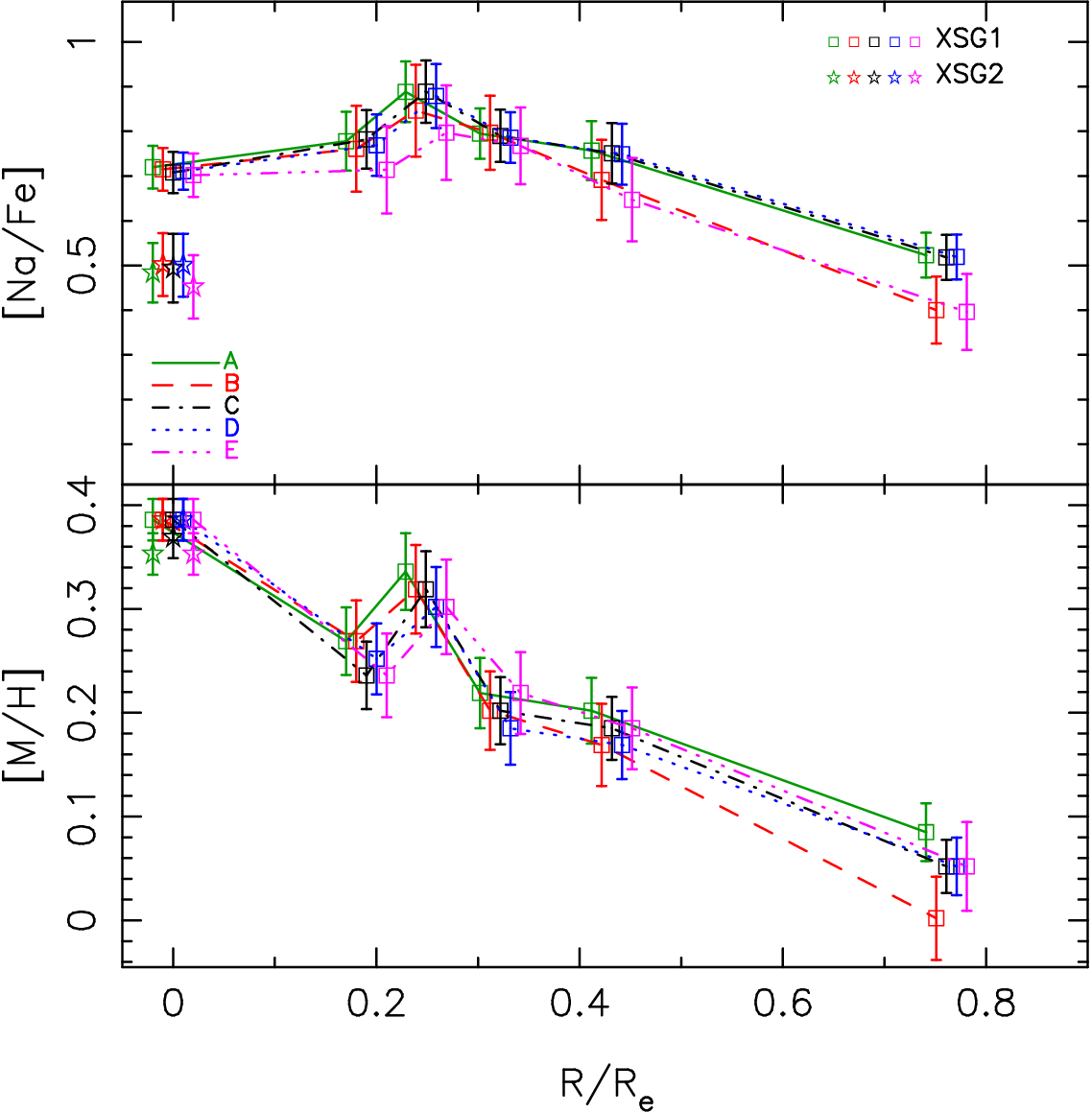}
\end{center}
\caption{{ Best-fit \nafe\ (upper-) and \zh\ (lower-panel) } for different radial bins of XSG1
  (squares) and the central spectrum of XSG2 (stars). The different
  fitting cases (see Sec.~\ref{sec:cases}) are plotted { with different colours and line types }
  (Tab.~\ref{tab:methods}). The error bars are shown at the 1\,$\sigma$ level.
  Notice that horizontal shifts have been applied among
  different fitting cases, to ease visualization.  A \nafe\ as high
  as 0.7--0.8\,dex is required to fit the data for XSG1, except for
  the outermost radial bin, where a lower \nafe\ value of
  0.4--0.5\,dex is found. The shape of the \nafe\ profile for XSG1
  is remarkably similar to that of \cfe\ and \afe\ (see figure~7 of
  LB16), but shifted to higher values. { XSG1 exhibits a negative metallicity gradient, typical
  of ETGs (see also LB16)}. The central spectrum of XSG2
  is characterized by lower Na abundance than XSG1, with \nafe=0.5$\pm$0.08\,dex.
  Notice that \nafe\ abundances should be taken
  as upper limits (see Sec.~\ref{sec:upperlimits}).  }
\label{fig:na_rad}
\end{figure}

{ The upper panel of } Fig.~\ref{fig:na_rad} shows the radial profile of \nafe\ in XSG1, and
the central value of XSG2, { while the bottom panel of the Figure shows the best-fitting values of 
\zh\ for all spectra}. Lines with different colours { and types} correspond to
the different fitting schemes considered.  All methods produce a flat
radial trend of \nafe\ in XSG1, with, on average, a high value of
$\sim$0.7--0.8\,dex, and a somewhat lower value of
$\sim$0.4--0.5\,dex in the outermost radial bin. We also
infer a high \nafe\ abundance ratio in XSG2 ($\sim$0.5\,dex), albeit
significantly lower than the central value for XSG1. Notice that among
all the fitting methods, the best-fit \nafe\ is consistent within
the 1\,$\sigma$ confidence level.  The high value of \nafe\ in XSG1
is consistent with the fact that this galaxy is the one with the highest
NaD line-strength among the SDSS ETGs with velocity dispersion
$\sigma\sim$300\,\kms\ analyzed by LB13 (i.e. \nai$\sim$5.5$\pm$0.15\,\AA\
within the SDSS fibre aperture).  On the other hand, XSG2 has a
significantly lower strength (\nai$\sim$4.85$\pm$0.2\,\AA),
consistent with the lower \nafe\ inferred here (see Tab.~\ref{tab:indices}).
Together with Fig.~\ref{fig:imf_rad_na}, Fig.~\ref{fig:na_rad} shows
that in order to fit all Na-sensitive features consistently in XSG1 and XSG2,
we need both a bottom-heavy IMF and \nafe\ overabundance.
Indeed, for \naii, \naiii, and \naiv, a bottom-heavy distribution
works as a catalyst to the response with \nafe, boosting the line
strengths significantly with respect to a standard IMF. Note
that for a Kroupa-like IMF, the central values of \naiii\ and \naiv
in XSG1 would imply \nafe$\gtrsim$1.2\,dex. This seems
rather unrealistic, even considering that \nafe\ can be as high as
0.8\,dex in the globular clusters of M31 \citep{Colucci:2014}.  Such
a high sodium abundance would imply \nai\ strengths around
10--11\,\AA\ (based on our models) -- far larger than the observed
values -- as well as extremely high metallicities, above 0.5\,dex (in
order to match the observed \mgfep\ in the innermost radial bin).  As
shown in Fig.~\ref{fig:imf_rad_na} and Fig.~\ref{fig:na_rad}, the
simultaneous fitting of the Na-sensitive line strengths rules out this
kind of scenario.
 
\subsection{Empirical response of Na-sensitive features to \afe}
\label{sec:aferesponses}
\begin{figure}
\begin{center}
\leavevmode
\includegraphics[width=8.2cm]{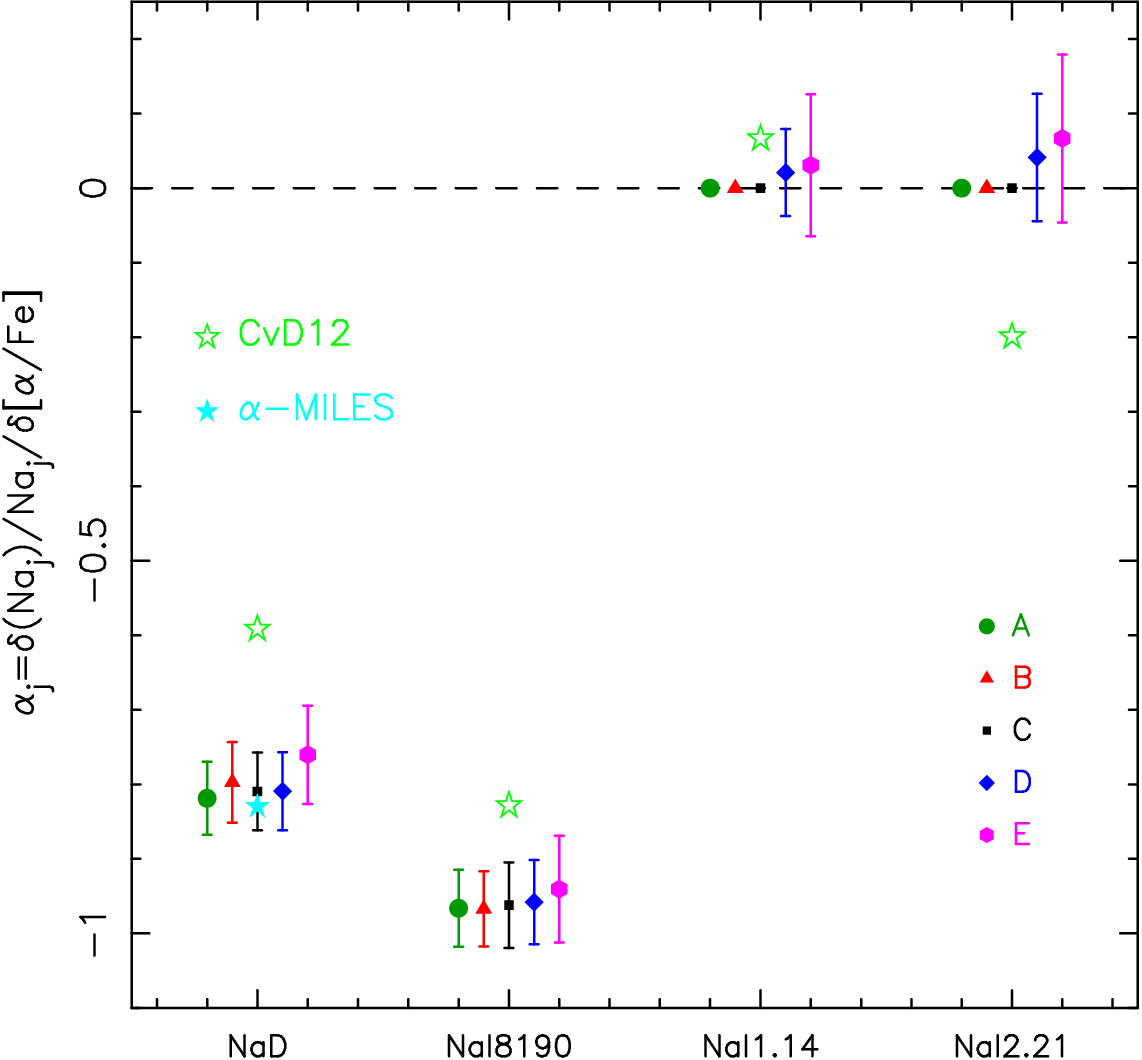}
\end{center}
\caption{Best-fit values of the responses of Na features, \{\alj\},
  to \afe\ (see Eqs.~\ref{eq:chi} and~\ref{eq:respal}).  From left to
  right, we plot values of \alj\ for \nai, \naii, \naiii, and \naiv ,
  respectively.  The different fitting cases (see
  Sec.~\ref{sec:cases}) are { colour (Tab.~\ref{tab:methods}) and symbol coded (see the legend 
  in the figure)}. The
  error bars are shown at the 1\,$\sigma$ level.  Notice that no error
  bars are shown when the \alj\ coefficients are kept fixed in the
  fitting.  Horizontal shifts have been applied among different
  fitting cases, for displaying reasons, { with shifts increasing from case A through E}.  
  For comparison, light--green
  { empty } stars show the values of \alj\ predicted for the CvD12 SSP models
  with an age of 13.5\,Gyr, solar metallicity, and a Chabrier
  IMF. Note that the CvD12 models are computed at fixed [Fe/H], rather
  than total metallicity (as in the extended--MILES models),
  complicating the comparison to our results, especially for \nai\ and
  \naii\ (which have a stronger dependence on \zh ). Remarkably, our
  best-fit results for \nai\ are fully consistent with (independent)
  model predictions based on $\alpha$--MILES models (cyan { filled} star; see
  the text and Tab.~\ref{tab:amilesresp}), taking the effect of
  \afe\ explicitly into account (at fixed \zh ).  }
\label{fig:aij}
\end{figure}

Fig.~\ref{fig:aij} shows the best-fit values of \alj, for all four
Na-sensitive lines, comparing them to the predictions from the CvD12
models (green stars), and the $\alpha$--MILES models (only for \nai,
cyan star; see Sec.~\ref{sec:indices} and Tab.~\ref{tab:amilesresp}).
We remark that the comparison to the CvD12 models should be taken with
caution, as these models are computed at fixed [Fe/H], whereas the Na--MILES,
$\alpha$--MILES, and extended--MILES models are all computed at fixed
total metallicity, \zh.  Interestingly, we find that the response of \naiii\ and
\naiv\ to \afe\ is consistent with zero (see magenta and blue circles,
with error bars, in the Figure). In the case of \naiii , this result
is fully consistent with the CvD12 prediction, while for \naiv, the
CvD12 models predict a slight anticorrelation with
\afe\ (i.e. $\rm\alpha_{NaI2.21}\sim -0.2$), not confirmed by our
analysis. For \nai\ and \naii , the CvD12 models lie systematically
above the best-fit values of $\rm\alpha_{NaI1.14}$ and
$\rm\alpha_{NaI2.21}$.  Qualitatively, this is expected, as both
\nai\ and \naii\ increase with {\sl total metallicity}, and thus the
\afe\ dependence is expected to be stronger at fixed \zh\ than at
fixed [Fe/H] (see Sec.~\ref{sec:assumptions}).
Note that the coefficient $\rm \alpha_{NaD}$ from the $\alpha$--MILES
models (cyan star), is more negative than the corresponding one from the CvD12
models. One of the most remarkable results of the present study is
that the best-fit value of $\rm \alpha_{NaD}$ is {\it fully consistent},
within the errors, with the prediction of the $\alpha$--MILES models. This
result is far from trivial, as in the $\alpha$--MILES models the
effect of \afe\ is taken into account by use of the recent [Mg/Fe]
determinations of MILES stars, along with theoretical corrections,
whereas in the present work, we fully rely on theoretical corrections for
the effect of \nafe, constraining, empirically, the effect of \afe.
The fact that the best-fit and $\alpha$--MILES values of $\rm\alpha_{NaD}$
are fully consistent, proves the internal consistency of
our modelling approach, and the robustness of these results.

\section{Caveats}
\label{sec:upperlimits}
An important aspect of the present analysis is that absolute values of
\nafe, as inferred from Na--MILES model predictions ({ see
upper-panel of } Fig.~\ref{fig:na_rad}), should be taken as upper limits.  Our model
line strengths depend mildly on how one treats the response of the
spectra of cool stars to \nafe. In our reference modelling approach
(see Sec.~\ref{sec:namiles}), we assume that the relative response of
stellar spectra to \nafe\ in stars with an effective temperature
$T_{\rm eff}\lesssim$3500\,K -- for which synthetic stellar
atmosphere models are not available -- is the same as for stars with
3500\,K (see Sec.~\ref{sec:models}). In App.~\ref{app:teff}, we
compare our reference model predictions (Fig.~\ref{fig:na_fits}) to
those from a linear extrapolation of the response to \nafe\ for stars
cooler than 3500\,K. The extrapolation does not affect significantly
\nai\ and \naii , while \naiii\ and \naiv\ increase slightly
($\lesssim$8\%), mostly at high metallicity and for a bottom-heavy
IMF. As a result, the use of temperature-extrapolated models results
in mildly lower values of \nafe\ (by $\sim$0.1\,dex) with respect to
the results shown in { the upper panel of  } Fig.~\ref{fig:na_rad}. Therefore, none of the conclusions in the
present work are affected. More importantly, one should note that our stellar
atmosphere models are based on LTE. For a solar-like
star, LTE  predicts weaker lines
\citep{Allende-Prieto03}, needing corrections that, for the { strongest and saturated}
lines, can be as high as an equivalent change in the abundance of
$\sim$0.5\,dex. For a lower temperature star (with $\rm T_{\rm eff} \sim 4000$~K), more relevant 
for models having old ages (such as those of our sample of ETGs), 
we may expect NLTE corrections in the range of 0.1--0.2~dex (based on figure~4 
of~\citealt{Lind11}).  
Hence, we may expect that
our LTE-based models predict weaker Na{\sc I} lines, leading to high
inferred values of \nafe, an issue that has to be addressed with future model
developments. We emphasize that our analysis matches \nai, \naii,
\naiii, and \naiv, simultaneously, suggesting that the non-LTE case
will likely affect all four indices in a similar way. If this
interpretation holds and non-LTE corrections were significant, the
Na-enhanced model grids in Fig.~\ref{fig:na_fits} should be affected
in two ways: (i) the values of \nafe\ (=\{0.3, 0.6, 0.9\}) for the
grids should be  lower than those reported in the Figure; and
(ii) as a consequence, the Na-enhanced grids would not shift so much 
towards lower values of \mgfep, with respect to Fig.~\ref{fig:na_fits}.
In order to test how the second effect would impact our results,
we re-ran the fitting case B (see Sec.~\ref{sec:cases}) by neglecting
the dependence of \mgfep\ on \nafe\ (i.e. we assume \mgfep\ values
from the extended--MILES models, with \nafe=0). Notice that in this
case, no extrapolation of the grids in Fig.~\ref{fig:na_fits} is
required, as the high values of \mgfep\ for the central
spectra of XSG1 and XSG2 can be roughly matched with the highest
metallicity ([M/H]=$+$0.22) of extended--MILES models. We found
that neglecting the dependence of \mgfep\ on \nafe\ produces an
IMF profile as well as \alj\ values (see Sec.~\ref{sec:aferesponses})
that are fully consistent with those for case B (Figs.~\ref{fig:imf_rad_na},
~\ref{fig:aij}), whereas the \nafe\ estimates would
(virtually) be $\sim$0.15\,dex higher than those in
Fig.~\ref{fig:na_fits}, hence not changing significantly our
conclusions.

\section{Discussion}
\label{sec:disc}

\subsection{Na abundance ratios in massive ETGs}
\label{sec:naetgs}
Based on independent stellar population models, \citet{Conroy:2014}
and \citet{Worthey:2014} reported \nafe$\sim$0.4~dex, in a set of stacked
ETG spectra from SDSS, over a range of velocity dispersion
$\sigma\sim$260--300\,\kms.  These values agree well with the estimate
for XSG2: \nafe=0.5$\pm$0.08\,dex.  \citet{Worthey:2014} report values
of \nafe\ as high as $\sim 0.55$~dex for individual Virgo cluster
galaxies, while \citet{CvD12b} find \nafe\ ratios as high as
$\sim$0.8\,dex in some ETGs, and even 1\,dex in the bulge of M31. Note
that XSG1 gives a similarly high value: \nafe$\sim$0.7\,dex.
Therefore, XSG1 features an ``extreme'' sodium abundance, as confirmed
by its NaD line strength, the highest among SDSS galaxies with similar
velocity dispersion (see Sec.~\ref{sec:results}). We have also modified our approach to 
fit the XSG1 and XSG2 line strengths simultaneously to \nai\ and \naii\ data 
for the 300\kms\ SDSS stacked spectrum of ETGs from LB13. In this case, we are able 
to match both the X-Shooter and SDSS data simultaneously, finding an abundance of 
\nafe$=0.42 \pm 0.05$~dex, fully consistent with \citet{Conroy:2014},
and an IMF slope of $\rm \Gamma_b = 2.7 \pm 0.4$ for the SDSS stack, consistent with the IMF--sigma relation 
derived by LB13 (see their figure~12).

It is important to emphasize that any estimate of \nafe\ should be actually taken as
an upper limit. Note we can compute the theoretical response of
stellar spectra to \nafe\ only for stars hotter than 3500\,K (see
Sec.~\ref{sec:namiles}). As discussed in App.~\ref{app:teff}, a linear
extrapolation of Na responses to cooler temperatures would likely lead
to lower \nafe\ estimates, by about 0.1\,dex. Moreover, as already
noticed by CvD12b, atomic Na{\sc I} transitions in the atmospheres of
late-type stars are prone to substantial departures from LTE
\citep{Bruls92, Gehren06, Andrievsky07, Lind11}. In a Sun-like star, LTE calculations predict 
weaker lines \citep{Allende-Prieto03} requiring corrections for the
strongest lines, which can be as high as an effective change in the abundance of
$\sim$0.5\,dex.  
For a lower temperature star (with $\rm T_{\rm eff} \sim 4000$~K), more relevant 
for models having old ages (as those of our sample of ETGs), 
we may expect non-LTE (NLTE) corrections in the range of 0.1--0.2~dex (based on figure~4 
of~\citealt{Lind11}). 
Hence, we may expect that NLTE models could enhance the predicted absorption in Na
indices, resulting in lower inferred values of \nafe. However, note that our LTE-based
methodology can match \nai, \naii, \naiii, and \naiv, simultaneously,
suggesting that NLTE corrections should be approximately the same for
all four Na lines -- an important aspect to test with future models.
We emphasize that although we have introduced four free
fitting parameters to match the four Na lines (the \alj\ constants;
see Eq.~\ref{eq:chi}), in practice the values of $\rm \alpha_{NaD}$,
$\rm \alpha_{NaI1.14}$ and $\rm \alpha_{NaI2.21}$ are fully consistent
with those of $\alpha$--MILES, and CvD12 models. Effectively, we 
are able to fit the four Na-sensitive line strengths of the seven X-Shooter
spectra -- spanning a range of age, metallicity, and \afe -- based on
only one ``extra'' free-fitting parameter (i.e. the $\rm
\alpha_{NaI8190}$).

We notice that previous works constrained \nafe\ only with one, or at
most two, Na-sensitive lines (\nai\ and \naii). For instance,
\citet{CvD12b} mostly rely on the \naii\ line, constraining sodium
abundance from its indirect effect on the spectrum, due to its
influence on the free electron abundance in the stellar atmospheres.
One important concern when using \nai\ is the effect of dust
absorption \citep{Sparks:1997, Jeong:2013}, either a foreground
component from our Galaxy (not the case given the redshift of our
sources) as well as intrinsic to the targeted systems.  However, the
observed \nai\ does not depend significantly on E(B-V) for massive
ETGs in SDSS (LB13). Moreover, the effect of reddening should be more
prominent in low-sigma (hence younger) galaxies, whereas, the inferred
\nafe\ is found to {\it increase} with velocity dispersion
\citep[based on \nad,][]{Conroy:2014}. Our work suggests a negligible
role of dust absorption on \nai , as this feature is matched
simultaneously to the NIR lines, where the effect of reddening is
completely negligible. This is not surprising, considering that our
targets have been intentionally selected from the pool of SDSS massive
galaxies with low internal extinction (see LB13).

Notwithstanding the above issues, a high sodium overabundance in
massive ETGs is a plausible scenario. However, an explanation 
is not clear.  High values of \nafe\ have been measured in 
individual stars in Globular Clusters (GCs). For instance,
\citet{Colucci:2014} found \nafe$\sim$0.8\,dex in M31 GCs.
In GCs, high Na abundances are usually associated with 
low oxygen abundances, i.e. the well-known Na--O anti-correlation
\citep[e.g.][]{Carretta:2009}. While one might be tempted to
relate this trend in GCs with the high [Na/Fe] in massive ETGs,
the observations do not support the existence of such anti-correlation
in galaxies.
For instance, Both \nafe\ and \ofe\ are found to increase with galaxy
velocity dispersion \citep{Conroy:2014}.


Regarding the chemical enrichment aspect of the problem, sodium can be
synthesized both in massive ($>$8\,M$_{\odot}$) and intermediate-mass
(3$<$M/M$_{\odot}<$8) stars. The former produce Na during the  hydrogen
burning process (via the $^{22}$Ne($p,\gamma$)$^{23}$Na reaction) and also via
carbon burning (where $^{23}$Na is an intermediate product of the full
$^{12}$C $+$ $^{12}$C chain). At solar metallicity, the stellar models
of \citet{Chieffi:2004} predict a sodium enrichment by a factor between
a few tens and $\sim$10$^3$ in the stellar mass range 13--35\,M$_{\odot}$.
The synthesized sodium is later ejected into the interstellar medium
through stellar winds and mass loss during the 
Wolf-Rayet phase, and subsequently via SN explosion.  Nucleosynthetic 
yields of Na from this channel are expected to increase significantly
with metallicity, even more than for other odd-Z elements (e.g. Al;
see~\citealt{Kobayashi:2006}).  Intermediate-mass stars, more massive
than $\sim$3\,M$_{\odot}$, contribute to the synthesis of Na in their
Asymptotic Giant Branch phase, where Na is produced via hot-bottom
burning (following the $^{22}$Ne($p,\gamma$)$^{23}$Na reaction) in their
convective envelope, and during their interpulse period. The stellar
models computed by \citet{Ventura:2013} show that the amount of Na
produced by massive AGB stars strongly increases with increasing
metallicity (up to the maximum value probed, i.e. $\sim 2 Z_\odot$) at
all stellar masses between 3 and 8\,M$_{\odot}$. At solar
metallicity the Na enrichment achieved by massive AGB stars is a
factor $\le$2. The freshly synthesized Na is released to the
interstellar medium via the heavy mass loss experienced by these stars
during their AGB phase and at the AGB tip.  The comparison of Na
yields computed for massive stars by \citet{Chieffi:2004} with
those calculated for massive AGB stars by \citet{Ventura:2013} 
indicates that massive stars are the dominant channel for  Na enrichment in
galaxies. Nevertheless, a bottom-heavy IMF, as measured in XSG1 and
XSG2, might make the contribution of massive AGB stars more relevant.
Notice that for both formation channels (SNe and AGB production), one
would expect a correlation of \nafe\ with metallicity, that we do not
see, as a function of galacto-centric distance.
\nafe\ does not show any significant radial variation in the inner
region of this galaxy (out to $\sim$0.5\,\re), in contrast to
the change in metallicity ($\sim$0.2\,dex; { see lower-panel of Fig.~\ref{fig:na_rad}}) over the same
radial range. Moreover, as mentioned in Sec.~\ref{sec:results}, the
radial behaviour of \nafe\ in XSG1 (Fig.~\ref{fig:na_rad}) is very
similar to \afe\ and \cfe\ (see figure~7 of LB16). A possible
explanation is that Na yields do rapidly reach a saturation at
super-solar metallicities, possibly giving no correlation of
\nafe\ with \zh\ in this regime.  Alternatively, under a
time-dependent IMF scenario \citep{Vazdekis1996, Vazdekis1997,
  Weidner2013, Ferreras2015}, both Mg and Na might have been produced
during the early, low-metallicity, top-heavy phase of galaxy formation.
In order to test these scenarios, a more detailed analysis of how Na is
produced in massive and AGB stars, at different metallicities, would
be required.


\subsection{The IMF of ETGs from Na lines}
Previous studies have claimed that in order to explain the optical
Na-sensitive lines of massive ETGs -- in particular \nai\ and \naii --
high sodium enhancement and a dwarf-rich IMF might be required
\citep{Spiniello:2012}.  The present work shows, for the first time in
a quantitative manner, that the combination of {\sl both} sodium
enhancement and a dwarf-rich IMF is crucial to explain the optical
{\it and} NIR Na-sensitive lines.

The interpretation of NIR Na lines has been hampered by a significant
mismatch between the observations and the models.  \citet[hereafter
  SAL15]{smith:2015} have recently shown that state-of-the-art (CvD12)
stellar population models struggle to reproduce the NIR \naiii\ in
most massive galaxies. { For a stacked spectrum of ETGs with $< \! \sigma \! > \approx 300$~\kms, 
SAL15 found that the observed line-strength of \naiii\ is significantly
stronger than that of a best-fitting model based on full spectral fitting. The mismatch
is $\sim$15\%, as shown by the blue cross and red circle in figure~3
of SAL15. Our results seem to reduce this discrepancy, as for
a bottom-heavy distribution (typical of the average population of massive ETGs), say \gammab=3.3, 
and for ``high'' \nafe\ ($\sim$0.6~dex),
our models predict \naiii\ strengths that are 15\% higher (i.e. similar to the
offset in figure~3 of SAL15) than what one would get when evaluating the
impact of \nafe\ on \naiii\ for a Kroupa-like IMF. 
Notice that this argument does not solve
the issue of an high \naiii\  for some massive galaxies with an IMF normalization close to Kroupa, as  found
by SAL15. However, it could help to reduce the tension.
In fact, even for a relatively ``light''
IMF, such as a bimodal IMF with \gammab=2.3 (corresponding to a mass-excess parameter, 
$\rm (M/L)/(M/L)_{Kroupa} \sim 1.3$, i.e. an $\rm M/L$ only slightly above that for a Kroupa distribution), 
the coupled effect of IMF and \nafe\ on Na features is not negligible (see upper--right panel of Fig.~\ref{fig:na_fits}): 
for \gammab=2.3 and \nafe$=0.9$~dex, one 
predicts an \naiii\ line-strength as high as for the case of a significantly bottom-heavy distribution (\gammab=3.3),
and lower \nafe ($\sim0.6$~dex).
Notice that this scenario would not work for a significantly heavier, single-segment, IMF, supporting
the conclusions of LB16.
}
Our work also shows that the way of implementing the effect of \nafe\ in
low-temperature stars will affect the NIR Na-sensitive line
strengths, especially in the high metallicity regime. However, the
mismatch becomes { significant } (although still below the 10\% level)
only for a bottom-heavy IMF.

The behaviour of the \naiv\ index is perhaps even more
complex. \citet{MenesesGoytia2015} found that both their and CvD12
models are unable to match the larger \naiv\ of massive ETGs (see also
R\"ock et al., in preparation).  \citet{ester:2009} reported a
significant correlation of \naiv\ with the optical \cfs\ index in
ETGs, suggesting a contribution of \cfe\ to the index that would
hamper its use to constrain IMF and \nafe.  Such a scenario is
disfavoured by our data, as XSG2 has significantly lower \cfe\ and
\afe\ than XSG1, but still a high \naiv\ line strength (see
Sec.~\ref{sec:sppars}). Moreover, in XSG1 \naiv\ and \naiii\ give
consistent \nafe, whereas the data for the central region of XSG2 is
where our models match less well the observations. Thus, at the
moment, the interpretation of \naiv\ remains incomplete, and needs
further investigations based on larger, homogeneous, samples.

The use of optical Na-sensitive lines to constrain the IMF has been
recently discussed by \citet[hereafter ZHD15]{Ziel:15}, based on new
(literature) \naii\ (\nai) data from the bulge of M31. The authors
argue that, because of the lower (higher) sensitivity of \naii\ (\nai)
to IMF in the extended-MILES, with respect to the CvD12, models, one
would infer a heavier IMF for the inner bulge of M31 with MILES,
relative to CvD12, models, inconsistent with dynamical
constraints \citep{Saglia:2010}. Results from the two sets of models
might be reconciled if either (i) \naii\ is more sensitive to \nafe ,
relative to \nai; or (ii) \nai\ decreases far more with \afe, relative
to \naii , in the MILES models. Indeed, neither of these two
hypotheses is supported by our analysis, as the ratio of the
responses to \afe\ (see Fig.~\ref{fig:aij}) and \nafe\ (see
Fig.~\ref{fig:na_fits}) are quite similar between the two sets of
models. However, one cannot constrain it in the regime of extreme 
\nafe\ overabundance, as might be the case for the bulge of M31
(for which CvD12 report a value of \nafe$\sim$1\,dex). Another
possibility is that the \nai\ line strengths used by ZHD15 -
originally observed by \citet{Davidge:1991, Davidge:1997} at much
lower resolution than \naii\ -- are, for some unknown reason, significantly
underestimated. Further, homogeneous, observations of all Na features
in the bulge of M31 (including \naiii\ and \naiv ), as well as the
computation of extended-MILES models where {\it both} \afe\ and
\nafe\ are varied {\it simultaneously}, would help to elucidate these
issues.

\section{Summary}
\label{sec:summary}
In the present work, we have performed, for the first time, a joint,
radially extended analysis of the four most prominent Na-sensitive
features in the optical and NIR spectral range 
{ Na \! D, Na \! I $\lambda$8190\AA, Na \! I $\lambda$1.14$\mu$m, and Na \! I $\lambda$2.21$\mu$m
(\nai, \naii, \naiii, and \naiv, respectively)}, in two massive, nearby, ETGs (named XSG1 and XSG2),
observed with the VLT/X-Shooter spectrograph at a high S/N across the
optical$+$NIR spectral window. The analysis relies on a modified version
of the extended--MILES stellar population models, covering the range
from 0.35 to 5\,$\mu$m, at moderately high, and uniform, spectral
resolution \citep{Vazdekis:12,RV:15}. By use of theoretical
responses of \nafe\ to empirical stellar spectra, we compute a
new set of (publicly available) population synthesis models, including a varying \nafe .  
For the first time it is possible to assess the effect of
\nafe\ on optical and NIR spectral features as a function
of (total) metallicity and IMF, allowing for a wide abundance range
(up to \nafe$\sim$1.2\,dex). Our main results are the following.
\begin{description}
 \item[- ] From a modelling point-of-view, we find that the effect of
   varying \nafe\ is strongly { coupled to} the IMF slope. For a
   bottom-heavy distribution, an increase of \nafe\ produces a
   stronger variation of Na line strengths than for a Kroupa-like
   IMF. In other words, a bottom-heavy IMF boosts the effect of \nafe,
   affecting in particular the NIR Na indices (\naii, \naiii, and \naiv ),
   as the impact of \afe\ increases with decreasing temperature for
   dwarf stars;
 \item[- ] For XSG1, where line strengths are measured at different
   galacto-centric distances, we find a significant negative radial
   gradient in all Na lines, although the error bars are larger for \naiv;
 \item[- ] For the first time, our models allow us to explain {\it
   simultaneously} all four Na-sensitive line strengths in massive ETGs, and in
   particular the large values observed in \naiii\ and \naiv ;
 \item[- ] To this effect, both a bottom-heavy IMF, as well as a
   ``high'' \nafe\ are required, in the central regions of both
   targets. The central best-fit value of \nafe\ for XSG1 (XSG2) is
   $\sim$0.7\,dex (0.5\,dex). In XSG1, the radial IMF trend from
   Na-sensitive features is consistent with the
   constraints from TiO features shown in our previous work (LB16),
   although with larger error bars in the outermost galaxy regions.
 \item[- ] The radial trend of \nafe\ in XSG1 is similar to 
   \afe\ and \cfe, featuring a constant value out to $\sim$0.5\,\re,
   followed by a lower \nafe\ (by $\sim$0.2\,dex) in the outermost radial bin
   ($\sim$0.8\,re).
 \item[- ] Considering the current uncertainties on theoretical atmosphere
   models (i.e. the lack of \nafe--enhanced models for cool stars, and
   LTE, rather than NLTE assumptions), all \nafe\ estimates should
   be taken as upper limits. For instance, modelling the
   response of cool stars to \nafe\ might lead to \nafe\ estimates
   mildly lower, by $\sim$0.1\,dex, than those reported here.
\end{description}
In conclusion, while there are still open issues related to the
interpretation of Na-sensitive absorption features in ETGs - requiring further
developments of stellar population models - the present work
represents a significant leap forward with respect to previous
studies, as for the first time we have shown that one can effectively 
match all prominent optical and NIR Na-sensitive features in massive
galaxies, provided that a varying IMF, and non-solar abundance ratios
are taken into account.

\section*{Acknowledgments}
Based on observations made with ESO Telescopes at the Paranal
Observatory under programmes ID 092.B-0378 and 094.B-0747 (PI: FLB).
FLB acknowledges the Instituto de Astrof\'isica de Canarias for the
kind hospitality when this project started.  We thank dr.  J. Alcal\'a
for the insightful discussions and help with the reduction of
X-Shooter spectra.  We also thank dr. F. D'Antona and dr. M. Salaris
for helpful discussions about Na production during different stages of
stellar evolution.
{ We thank the anonymous referee for his/her
useful comments that helped us to improve our manuscript.}  
We have made extensive use of the SDSS database
(http://www.sdss.org/collaboration/credits.html).  MC acknowledges
support from a Royal Society University Research Fellowship.  We
acknowledge support from grant AYA2013-48226-C3-1-P from the Spanish
Ministry of Economy and Competitiveness (MINECO).


\appendix

\section{Effect of low-temperature stars on \nafe--enhanced models}
\label{app:teff}
Since theoretical stellar spectra do not cover temperatures cooler
than \Teff=3500\,K (see Sec.~\ref{sec:namiles}), for stars with
\Teff$<$3500\,K in the empirical stellar library (see
Sec.~\ref{sec:models}), we applied differential corrections for
\nafe\ by assuming \Teff=3500\,K. This is a rather conservative
approach, as Na-sensitive absorption lines strengthen with
decreasing temperature. Hence, our \nafe-enhanced SSP spectra might
underestimate the sensitivity of Na indices to \nafe , especially for
a bottom-heavy IMF. To assess the impact of \Teff\ on our results, we
compute a set of models for which we linearly {\sl extrapolate}
the \nafe\ differential corrections to \Teff<3500\,K.
For each stellar spectrum, given \Feh\ and $\rm \log g$,
we use our interpolation algorithm (see
Sec.~\ref{sec:models}) to compute differential corrections (at each
wavelength in the spectrum) in the temperature range from 3500\,K to
4500\,K, for the chosen \Feh\ and $\rm \log g$. The differential
correction is linearly extrapolated, with respect to \Teff, at the desired
temperature value.  We compute the \nafe-enhanced SSP models by 
exchanging the input stellar library spectra by the corresponding
\nafe-enhanced ones, i.e. exactly in the same way as for the base
model SSPs.  We finally assemble the SSP spectra as described in
\citet{RV:16}.

\begin{figure*}
\begin{center}
\leavevmode
\includegraphics[width=15cm]{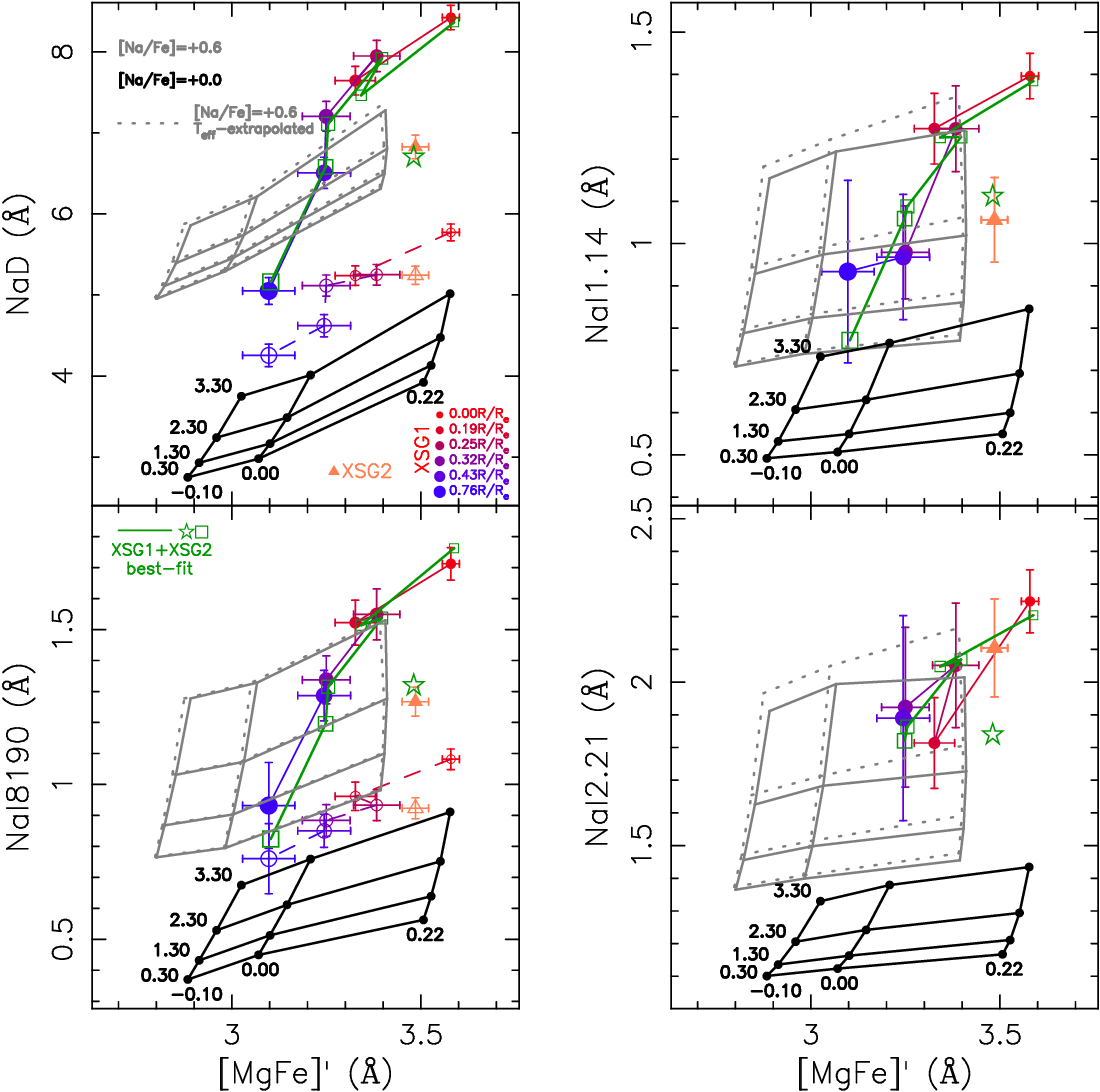}
\end{center}
\caption{Same as Fig.~\ref{fig:na_fits}, but plotting only models with
  \nafe$=0$ (black grids) and \nafe$=0.6$~dex (grey grids),
  respectively, with dotted grids showing the case where in the
  low-temperature regime ($<3500$~K) the effect of Na abundance on
  stellar atmospheres is linearly extrapolated, rather than assumed to
  be the same as for $\rm Teff=3500$~K.  Notice that the effect of
  Teff is almost negligible for \nai\ and \naii , while for
  \naiii\ and \naiv\ the extrapolation increases the model line
  strengths, in particular for models with high metallicity, and
  bottom-heavy IMF. As a result, one would infer somewhat lower Na
  abundances (by $\sim 0.1$~dex) than those reported in
  { the upper panel of } Fig.~\ref{fig:na_rad}.  }
\label{fig:na_fits_teff}
\end{figure*}

Fig.~\ref{fig:na_fits_teff} compares Na--MILES model predictions with
those of \Teff--extrapolated models, for \nafe=0.6\,dex. As expected,
the effect of the extrapolation is to increase the Na-sensitive line
strengths. The variation is very small for \nai\ and \naii, while it
is somewhat larger, albeit still small, for the NIR \naiii\ and
\naiv\ indices. Notice that the extrapolation affects more the
high-metallicity (i.e. lower-temperature) regime, as well as models
with a bottom-heavy IMF.  Since the corresponding variation of line
strengths is lower than $\sim$10\%, fitting observed line strengths
with \Teff--extrapolated, rather than our reference, Na--MILES models
give very similar results to those presented in Sec.\ref{sec:results}.
The most noticeable difference is that for \Teff--extrapolated models
we infer mildly lower (by $\sim$0.1\,dex) \nafe\ estimates than
those in Fig.~\ref{fig:na_rad}, not affecting any of the conclusions
of our work.

\section{Comparison of extended--MILES and CvD12 model predictions}
\label{app:rescaled}
Fig.~\ref{fig:na_imf_rescaled} compares predictions of Na-sensitive
line strengths for CvD12 and extended--MILES stellar population
models.  The Figure is similar to Fig.~\ref{fig:na_fits}, showing the
Na line strengths as a function of \mgfep . However, since CvD12
models include only SSPs with either Chabrier or unimodal
(i.e. single-segment) IMF, Fig.~\ref{fig:na_imf_rescaled} plots the
extended--MILES models with varying unimodal, rather than bimodal, IMF
slope, with 0.3$\le\Gamma\le$2.3. For CvD12 models, we plot
predictions for either a Chabrier or a unimodal, bottom-heavy, IMF
with $\Gamma$=2.3 (see grey stars in the Figure). As far as the
dwarf-to-giant ratio or the stellar mass-to-light ratio are concerned,
the case of a unimodal IMF with slope $\Gamma\sim$0.8 roughly
corresponds to a MW-like distribution (see, e.g., F13 and LB13).
Hence, the predictions of CvD12 models for a Chabrier distribution can
be compared to those for extended--MILES models with $\Gamma\sim$0.8.

\begin{figure*}
\begin{center}
\leavevmode
\includegraphics[width=14cm]{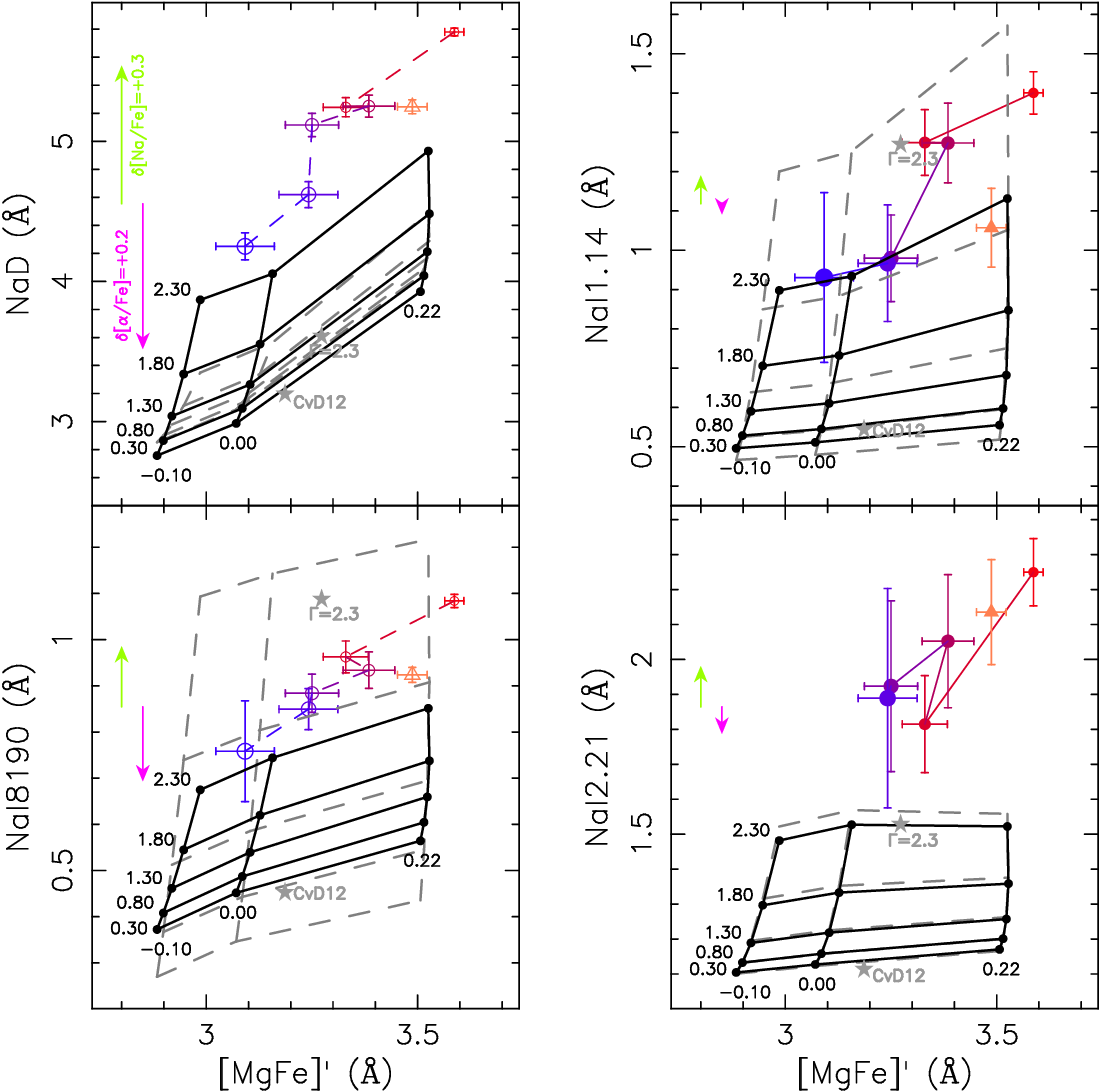}
\end{center}
\caption{ Same as Fig.~\ref{fig:na_fits}, but plotting predictions for
  unimodal, rather than bimodal, extended--MILES SSPs. The \nai\ and
  \naii\ line strengths are not corrected to \afe$=0$ (see filled
  circles and triangles in the left panels of Fig.~\ref{fig:na_fits}).
  Black grids correspond to models with \nafe$=0$, varying metallicity
  (\zh$\rm =-0.1, 0, +0.22$) and unimodal IMF slope ($\Gamma=+0.3,
  0.80, 1.30, 1.80, 2.30$), for an age of $11.2$~Gyr. Notice that $\Gamma =
  1.35$ corresponds to the Salpeter IMF, while $\Gamma \sim 0.8$
  roughly matches the mass-to-light ratio for a Kroupa-like IMF.  To
  make the figure more clear, only models with \nafe$=0$ are plotted.
  Grey grids are extended--MILES model predictions rescaled to match
  the IMF response of CvD12 models (see the text).  Grey stars are
  predictions from publicly available CvD12 SSP models, with solar
  metallicity, age of $13.5$~Gyr, and Chabrier and unimodal
  bottom-heavy ($\Gamma=2.3$) IMFs, respectively.  In each panel,
  light--green and magenta arrows show the effect of increasing
  \afe\ (\nafe) by $+0.2$~dex ($+0.3$~dex) in CvD12 models.  }
\label{fig:na_un}
\end{figure*}

\begin{figure}
\begin{center}
\leavevmode
\includegraphics[width=8.5cm]{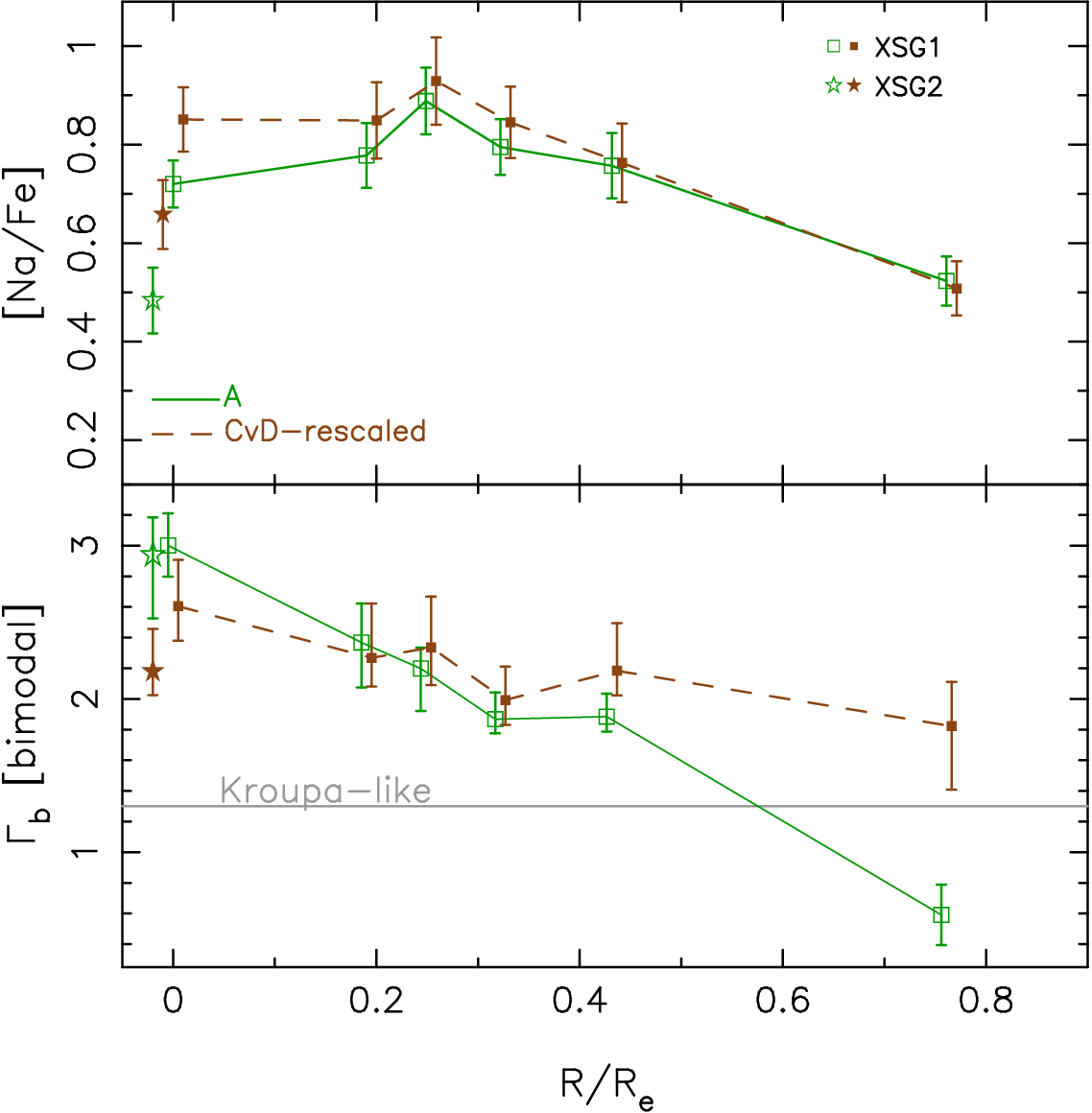}
\end{center}
\caption{\nafe\ ratios (top) and IMF slopes (bottom) as a function of
  galacto-centric distance, for XSG1 (squares) and XSG2 (stars),
  respectively. Green, { empty}, symbols are the same as in { the upper panel of} Fig.~\ref{fig:na_rad}
  and Fig.~\ref{fig:imf_rad_na}, respectively. Brown, { filled}, symbols, { connected by a dashed} line,
  are { results } obtained by fitting observed line strengths with
  Na--MILES models rescaled to match the response of CvD12 models to
  IMF and \nafe\ (see the text). Notice that this approach is only
  meant to test the robustness of our results against uncertainties on
  state-of-the-art stellar population models.  }
\label{fig:na_imf_rescaled}
\end{figure}

Fig.~\ref{fig:na_imf_rescaled} shows that the sensitivity of Na lines
to the IMF differs significantly between CvD12 and extended--MILES
models. The \nai\ (\naii ) is more (less) sensitive to IMF in
extended--MILES, than CvD12, models, in agreement with
\citet{spiniello:2015} and~\citet{Ziel:15}. The difference is even
more pronounced for \naiii\ (see~\citealt{smith:2015}), where CvD12
features a stronger sensitivity than the extended--MILES models. For
\naiv , predictions from different models are consistent, though the
effect of IMF variations is slightly larger for CvD12 than
extended--MILES models. The origin of such discrepancy is still
unclear, but mostly arises from the way different models attach stars
to the isochrones, while other ingredients (e.g. the adopted
isochrones) will likely play a secondary role
(see~\citealt{spiniello:2015}).  For the purpose of the present work,
we want to test the impact of differences in the sensitivity of the
models to the IMF on the derivation of the IMF slopes and
\nafe\ abundance ratios in XSG1 and XSG2.  To this effect, we have
first rescaled Na-MILES models to match the IMF response of CvD12
models.  For both sets of models, we compute the IMF sensitivity of Na
lines as follows:
\begin{equation}
\rm 
 \gamma_{Na_j} =  Na_j(\Gamma=2.3)-Na_{j,ref} ,
\label{eq:respimf}
 \end{equation}
where the index $\rm j$ defines a given Na line (see
Sec.~\ref{sec:fitting}), $\rm Na_{j,ref}$ are line strengths computed
for a MW-like IMF, i.e. a Chabri\'er (Kroupa) distribution for CvD12
(MILES) models.  For both sets of models, $\rm Na_j(\Gamma=2.3)$ are
line strengths of an SSP model having unimodal, bottom-heavy, IMF with
slope $\Gamma$=2.3. All line strengths in Eq.~\ref{eq:respimf} are
computed for SSP models with solar metallicity and an (old) age of
$\sim$13\,Gyr.  The values of \gaj , as well as the {\it relative}
responses of Na lines to IMF (defined as $\rm \gamma_j/Na_{j,ref}$)
are reported in Tab.~\ref{tab:imfresp}.
\begin{table}
\centering
\small
 \caption{Response of Na features to IMF (see Eq.~\ref{eq:respimf}) for CvD12 and extended--MILES 
 SSP models (cols.2 and~3, respectively). Values in parentheses are relative responses, defined 
 as $\rm \gamma_j/Na_{j,ref}$ (see the text).}
  \begin{tabular}{c|c|c}
   \hline
 Index  &      \gaj$\rm _{CvD}$      &      \gaj$\rm _{MILES}$  \\
       &          $\AA$              &               $\AA$       \\
 (1)   &            (2)             &             (3)              \\
   \hline
 \nai     &  0.41(0.13) &  0.92(0.28) \\
 \naii    &  0.63(1.40) &  0.23(0.43) \\ 
 \naiii   &  0.73(1.34) &  0.40(0.73) \\ 
 \naiv    &  0.41(0.37) &  0.37(0.32) \\ 
  \hline
  \end{tabular}
\label{tab:imfresp}
\end{table}
For each value of \nafe\ ($0, 0.3, 0.6, 0.9, 1.2$, see
Sec-~\ref{sec:models}), we have then rescaled Na-MILES models, for
bimodal distributions, as follows:
\begin{equation}
\rm 
 Na^{\prime}_j(\Gamma_b) = Na_{j,ref} + \frac{\gamma_{j,CvD}}{\gamma_{j,MILES}} \cdot \left[ Na_j(\Gamma_b) - Na_{j,ref} \right]
\label{eq:milesp}
\end{equation}
The rescaled models are shown as grey dashed grids in
Fig.~\ref{fig:na_un}, for the case with \nafe$=0$. As expected, the
effect is to increase (decrease) significantly the IMF sensitivity of
\naii\ and \naiii\ (\nai ), while leaving the \naiv\ grid almost
unaltered.  Using similar equations as \ref{eq:respimf}
and~\ref{eq:milesp}, we further modified the IMF-rescaled Na-MILES
models in order to match also the sensitivity of CvD12 models to
\nafe.  Hereafter, we refer to these modified models as CvD-matched
(Na-MILES) models.

We emphasize that, for the purpose of the present work, we do not aim
at discussing the origin of differences among Na--MILES and CvD12
models, but rather to test the possible impact of such differences on
our results. Fig.~\ref{fig:na_imf_rescaled} compares best-fit IMF
slopes and \nafe\ ratios obtained with Na--MILES models (green curves
in the Figure; see case A in Tab.~\ref{tab:methods}) to those obtained
for CvD-matched models (brown curves in the Figure). In the latter
case, we have applied no IMF constraint for XSG1 (see
Sec.~\ref{sec:cases}). Interestingly, we get similar results from
Na-MILES and CvD-matched models, in that we need high \nafe\ ratios,
as well as a bottom-heavy IMF in the innermost radial bins, as well as
an IMF radial gradient for XSG1, to fit our data.  However, in the
four inner radial bins of XSG1, as well as for the central bin of
XSG2, we tend to infer higher \nafe , and lower \gammab , for
CvD-matched than Na-MILES models.  For the innermost bin of XSG1
(XSG2), \nafe\ is $\sim 0.15$~dex ($\sim 0.2$~dex) higher, while
\gammab\ is $\sim 0.4$ ($\sim 0.7$) lower, for CvD12-matched than
Na-MILES.  In the outermost bins of XSG1, results are similar for both
sets of models, with lower \nafe\, and a lighter IMF
slope~\footnote{Notice that when no IMF constraints are imposed for
  XSG1, we get significantly larger error bars on IMF slope for
  Na-MILES (see cases B and E in Fig.~\ref{fig:imf_rad_na}) than
  CvD-matched models (brown curve in Fig.~\ref{fig:na_imf_rescaled}),
  in particular for the outermost radial bins. This happens as for
  CvD-matched models, \naii\ and \naiii\ have significantly larger
  sensitivity to IMF than \nafe , while for Na-MILES, all Na features
  have significant sensitivity to both IMF and \nafe\ (see
  Sec.~\ref{sec:modelsbehaviour}), implying a stronger degeneracy
  between IMF and \nafe\ (hence larger error bars) in the latter case.
}, than in the innermost radial bins. In other words, CvD12-matched
models give a shallower IMF (steeper \nafe ) trend than Na-MILES
models, with a somewhat lighter, yet bottom-heavy, IMF slope in the
innermost radial bins of both our targets.  Notice that, in a somewhat
similar way to these results, \citet{spiniello:2015} found that
CvD12 and MILES models tend to predict similar trends of IMF slope
with velocity dispersion in ETGs, though the MILES models favour slightly
higher slopes for the most massive ETGs.  In any case, the main point
for the present work is that our results are robust against
uncertainties of state-of-the-art stellar population models on the IMF
response of Na-sensitive features.

\begin{thebibliography}{99}

\bibitem[Allende Prieto, Hubeny, Lambert(2003)]{Allende-Prieto03}
Allende Prieto, C., Hubeny, I., Lambert, D.~L., 2003, ApJ, 591, 1192

\bibitem[\protect\citeauthoryear{Allende Prieto et~al.}{2008}]{Allende-Prieto08}
Allende Prieto, C., et al. 2008, AJ, 136, 2070
  
\bibitem[\protect\citeauthoryear{Alonso, Arribas \& Mart\'{\i}nez-Roger}{Alonso et~al.}{1995}]{Alonso95}
Alonso, A., Arribas, S., Mart\'{\i}nez-Roger, C., 1995, \aap, 297, 197

\bibitem[\protect\citeauthoryear{Alonso, Arribas \& Mart\'{\i}nez-Roger}{Alonso et~al.}{1996}]{Alonso96}
Alonso, A., Arribas, S., Mart\'{\i}nez-Roger, C., 1996, \aap, 313, 873

\bibitem[\protect\citeauthoryear{Alonso, Arribas \& Mart\'{\i}nez-Roger}{Alonso et~al.}{1999}]{Alonso99}
Alonso, A., Arribas, S., Mart\'{\i}nez-Roger, C., 1999, \aaps, 140, 261

\bibitem[Andrievsky al.(2006)]{Andrievsky07} Andrievsky, S.~M., Spite, M., Korotin, S.~A., Cayrel, R., 
Hill, V., Fran\c ois, P. 2007, A\&A, 464, 1081

\bibitem[Asplund et al.(2009)]{Asplund09}
Asplund, M., Grevesse, N., Sauval, A.~J., Scott, P., 2009, ARA\&A, 47, 481

\bibitem[\protect\citeauthoryear{Barklem et al.}{2000}]{Barklem00}
Barklem P.~S., Piskunov N., O'Mara B.~J., 2000, A\&AS, 142, 467
  
\bibitem[\protect\citeauthoryear{Bautista}{1997}]{Bautista97}
Bautista M.~A., 1997, A\&AS, 122,  

\bibitem[Bruls et al.(1992)]{Bruls92}
Bruls, J.~H.~M.~J., Rutten, R.~J., Shchukina, N.~G., 1992, A\&A, 265, 237

\bibitem[\protect\citeauthoryear{Cappellari \& Emsellem}{2004}]{Cap:2004}
Cappellari, M., Emsellem, E., 2004, PASP, 116, 138

\bibitem[\protect\citeauthoryear{Cappellari et al.}{2012}]{Capp:12}
Cappellari M., et al., 2012, Natur, 484, 485 

\bibitem[\protect\citeauthoryear{Cappellari et al.}{2013}]{Capp:13}
Cappellari M., et al., 2013, MNRAS, 432, 1862 

\bibitem[Carretta et al.(2009)]{Carretta:2009}
Carretta, E., Bragaglia, A., Gratton, R., Lucatello, S., 2009, A\&A, 505, 139C

\bibitem[\protect\citeauthoryear{Cassisi et~al.}{2000}]{Cassisi00}
Cassisi, S., Castellani, V., Ciarcelluti, P., Piotto, G., Zoccali, M., 2000, \mnras, 315, 679

\bibitem[\protect\citeauthoryear{Cassisi et~al.}{2004}]{Cassisietal04}
Cassisi, S., Salaris, M., Castelli, F., Pietrinferni, A., 2004, \apj, 616, 498

\bibitem[Castelli \& Kurucz(2004)]{CastelliKurucz04}
Castelli, F., Kurucz, R. L. 2004, arXiv:astro-ph/0405087

\bibitem[\protect\citeauthoryear{Cenarro et~al.}{2001a}]{CATI}
Cenarro, A.~J., Cardiel, N., Gorgas, J., Peletier, R.~F., Vazdekis, A., Prada, F., 2001a, \mnras, 326, 959

\bibitem[Cenarro et al. (2003)]{Cenarro:2003}
Cenarro, A.~J., Gorgas, J., Vazdekis, A., Cardiel, N., Peletier, R.~F., 2003, MNRAS, 339, L12

\bibitem[\protect\citeauthoryear{Cenarro et~al.}{2007a}]{MILESII}
Cenarro, A.~J., et~al., 2007a, \mnras, 374, 664

\bibitem[\protect\citeauthoryear{Cervantes \& Vazdekis}{2009}]{CV09} 
Cervantes, J. L., Vazdekis, A., 2009, MNRAS, 392, 691

\bibitem[\protect\citeauthoryear{Chabrier}{2003}]{Chabrier:03}
Chabrier G., 2003, PASP, 115, 763

\bibitem[\protect\citeauthoryear{Chabrier, Hennebelle, \& Charlot}{2014}]{Chabrier:14} 
Chabrier G., Hennebelle P., Charlot S., 2014, ApJ, 796, 75 

\bibitem[Chieffi \& Limongi(2004)]{Chieffi:2004}
Chieffi, A., Limongi, M., 2004, ApJ, 608, 405

\bibitem[\protect\citeauthoryear{Coelho et~al.}{2005}]{Coelho05}
Coelho, P., Barbuy, B., Melendez, J., Schiavon, R., Castilho, B.,  2005, \aap,  443, 735

\bibitem[\protect\citeauthoryear{Coelho et~al.}{2007}]{Coelho07}
Coelho, P., Bruzual, G., Charlot, S., Weiss, A., Barbuy, B., Ferguson, J.~W., 2007, \mnras, 382, 498

\bibitem[Colucci, Bernstein, Cohen(2014)]{Colucci:2014}
Colucci, J.~E., Bernstein, R.~A., Cohen, J., 2014, ApJ, 797, 116

\bibitem[\protect\citeauthoryear{Conroy \& van Dokkum}{2012a}]{CvD12a}  
Conroy, C., van Dokkum, P., 2012a, ApJ, 747, 69 (CvD12)

\bibitem[\protect\citeauthoryear{Conroy \& van Dokkum}{2012b}]{CvD12b}  
Conroy, C., van Dokkum, P., 2012b, ApJ, 760, 71

\bibitem[Conroy, Graves, van Dokkum(2014)]{Conroy:2014}
Conroy, C., Graves, G.~J., van Dokkum, P.~G., 2014, ApJ, 780, 33C

\bibitem[\protect\citeauthoryear{Cunto et al.}{1993}]{Cunto93}
Cunto W., Mendoza C., Ochsenbein F., Zeippen C.~J., 1993, A\&A, 275, L5 

\bibitem[Cushing et al.(2005)]{IRTFI}
Cushing, M.~C., Rayner, J.~T., Vacca, W.~D., 2005, ApJ, 623, 1115

\bibitem[Davidge(1991)]{Davidge:1991}
Davidge, T.~J., 1991, AJ, 101, 884

\bibitem[Davidge(1997)]{Davidge:1997} Davidge, T.~J., 1997, AJ, 113, 985

\bibitem[\protect\citeauthoryear{Ferreras  et al.}{2013}]{Ferr:13}  
Ferreras, I., La Barbera, F., de la Rosa, I.~G., Vazdekis, A., de
Carvalho, R.~R., Falc\'on-Barroso, J., Ricciardelli, E., 2013, MNRAS,
429, L15 (FER13)

\bibitem[Ferreras et al.(2015)]{Ferreras2015}
Ferreras, I., Weidner, C., Vazdekis, A., La Barbera, F., 2015, MNRAS, 448, 82

\bibitem[Girardi et al.(2000)]{Padova00}
Girardi, L., Bressan, A., Bertelli, G., Chiosi, C., 2000, A\&AS, 141, 371

\bibitem[Gehren et al.(2006)]{Gehren06}
Gehren, T., Shi, J.~R., Zhang, H.~W., Zhao, G., Korn, A.~J., 2006, A\&A, 451,

\bibitem[Grevesse \& Sauval(1998)]{GrevesseSauval98}
Grevesse, N., Sauval, A.~J., 1998, SSrv, 85, 161

\bibitem[\protect\citeauthoryear{Hopkins}{2013}]{Hopkins:13} 
Hopkins P.~F., 2013, MNRAS, 433, 170

\bibitem[Irwin(1981)]{Irwin81}
Irwin, A.~W., ApJS, 45, 621

\bibitem[Jeong et al.(2013)]{Jeong:2013}
Jeong, H., Yi, S.~K., Kyeong, J., Sarzi, M., Sung, E.-C., Oh, K., 2013, ApJS, 208, 7J

\bibitem[Kobayashi et al.(2006)]{Kobayashi:2006}
Kobayashi, C., Umeda, H., Nomoto, K., Tominaga, N., Ohkubo, T., 2006, ApJ, 653, 1145K

\bibitem[Koesterke, Allende Prieto, Lambert(2008)]{Koesterke08}
Koesterke, L., Allende Prieto, C., Lambert, D.~L., 2008, ApJ, 680, 764

\bibitem[Koesterke(2009)]{Koesterke09}
Koesterke, L., 2009, AIPC, 1171, 73

\bibitem[\protect\citeauthoryear{Kroupa}{2001}]{Kroupa:01}
Kroupa P., 2001, MNRAS, 322, 231


\bibitem[\protect\citeauthoryear{La Barbera et al.}{2008}]{LBdC08}
La Barbera, F., et al. 2008, PASP, 120, 681


\bibitem[\protect\citeauthoryear{La Barbera et al.}{2010a}]{SpiderI}  
La Barbera, F., de Carvalho, R.~R., de la Rosa, I.~G., Lopes, P.~A.~A.,
Kohl-Moreira, J.~L., Capelato, H.~V., 2010a, MNRAS, 408, 1313

\bibitem[\protect\citeauthoryear{La Barbera et al.}{2012}]{LB:12}
La Barbera F., Ferreras I., de Carvalho R.~R., Bruzual G., 
Charlot S., Pasquali A., Merlin E., 2012, MNRAS, 426, 2300 

\bibitem[\protect\citeauthoryear{La Barbera et al.}{2013}]{LB:13} 
La Barbera, F., Ferreras, I., Vazdekis, A., de la Rosa, I.~G., de
Carvalho, R.~R., Trevisan, M., Falc\'on-Barroso, J., Ricciardelli, E.,
2013, MNRAS, 433, 3017 (LB13)

\bibitem[\protect\citeauthoryear{La Barbera, Ferreras, \& Vazdekis}{2015}]
{LB:15} La Barbera F., Ferreras I., Vazdekis A., 2015, MNRAS, 449, L137 

\bibitem[\protect\citeauthoryear{La Barbera et al.}{2016}]{LB:16}
La Barbera F., Vazdekis A., Ferreras I., Pasquali A., Cappellari M.,
Mart\'\i n-Navarro I., Sch\"onebeck F., Falc\'on-Barroso J., 2016,
MNRAS, 457, 1468

\bibitem[\protect\citeauthoryear{Leier et al.}{2015}]{Leier:15} 
Leier D., Ferreras I., Saha P., Charlot S., Bruzual G., La Barbera F., 
2015, preprint arXiv:1512.00462

\bibitem[Lind et al.(2011)]{Lind11} Lind, K., Asplund, M., Barklem, P.~S., Belyaev, A.~K., 2011, A\&A, 
528, 103

\bibitem[\protect\citeauthoryear{Maraston}{2005}]{Maraston05}
Maraston, C., 2005, \mnras, 362, 799

\bibitem[\protect\citeauthoryear{Marigo et~al.}{2008}]{Marigo08}
{{Marigo}, P., {Girardi}, L., {Bressan}, A., {Groenewegen}, M.~A.~T., {Silva}, L., {Granato}, G.~L.}, 2008, \aap, 482, 883

\bibitem[\protect\citeauthoryear{M\'armol-Queralt\'o et al.}{2009}]{ester:2009} 
M\'armol-Queralt\'o, E., Cardiel, N., S\'anchez-Bl\'azquez, P., Trager, S.~C., 
Peletier, R.~F., Kuntschner, H., Silva, D.~R., Cenarro, A.~J., Vazdekis, A., 
Gorgas, J., 2009, ApJ, 705, 199

\bibitem[\protect\citeauthoryear{Mart{\'{\i}}n-Navarro et al.}{2015a}]{NMN:15a}
Mart{\'{\i}}n-Navarro I., La Barbera F., Vazdekis A.,
Falc{\'o}n-Barroso J., Ferreras I., 2015a, MNRAS, 447, 1033

\bibitem[\protect\citeauthoryear{Mart{\'{\i}}n-Navarro et al.}{2015b}]{NMN:15b}
Mart{\'{\i}}n-Navarro I., et al., 2015b, ApJ, 806, L31 

\bibitem[\protect\citeauthoryear{Mart{\'{\i}}n-Navarro et al.}{2015c}]{NMN:15c}
Mart{\'{\i}}n-Navarro I., La Barbera F., Vazdekis A., 
Ferr{\'e}-Mateu A., Trujillo I., Beasley M.~A., 2015c, MNRAS, 
451, 1081 

\bibitem[\protect\citeauthoryear{McConnell et al.}{2015}]{McConnell:15}
McConnell N.~J., Lu J.~R., Mann A.~W., 2015, preprint arXiv:1506.07880 

\bibitem[Meneses-Goytia et al.(2015)]{MenesesGoytia2015}
Meneses-Goytia, S., Peletier, R.~F., Trager, S.~C., Vazdekis, A., 2015, A\&A, 582, A97

\bibitem[\protect\citeauthoryear{M\'esz\'aros et al.}{2012}]{Meszaros:12}
M\'esz\'aros, Sz., Allende Prieto, C., Edvardsson, B., Castelli, F., Garc\'\i a P\'erez, A. E.,
Gustafsson, B.,  Majewski, S.~R., Plez, B., Schiavon, R., Shetrone, M.,
de Vicente, A., 2012, AJ, 144, 120

\bibitem[Modigliani et al.(2010)]{Mod:2010}
Modigliani, A., Goldoni, P., Royer, F., et al. 2010, in Society of
Photo-Optical Instrumentation Engineers (SPIE)
Conference Series, Vol.~7737, Observatory Operations:
Strategies, Processes, and Systems III, ed. D. R. Silva, A. B. Peck, \& T. Soifer

\bibitem[\protect\citeauthoryear{Nahar}{1995}]{Nahar95}
  Nahar S.~N., 1995, A\&A, 293,  

\bibitem[Navarro-Gonz\'alez et al.(2013)]{NavarroGonzalez:2013}
Navarro-Gonz\'alez, J., Ricciardelli, E., Quilis, V., Vazdekis, A., 2013, MNRAS, 436, 3507

\bibitem[Nomoto et al.(2006)]{Nomoto:2006}
Nomoto, K., Tominaga, N., Umeda, H., Kobayashi, C., Maeda, K., 
2006, Nuclear Physics, 777, 424 

\bibitem[\protect\citeauthoryear{Oser et al.}{2010}]{Oser:10} 
Oser L., Ostriker J.~P., Naab T., Johansson P.~H., Burkert A., 2010, ApJ, 
725, 2312

\bibitem[\protect\citeauthoryear{Padoan \& Nordlund}{2002}]{PadNor:02}
Padoan P., Nordlund {\AA}., 2002, ApJ, 576, 870 

\bibitem[\protect\citeauthoryear{Pietrinferni et~al.}{2004}]{Pietrinferni04}
Pietrinferni, A., Cassisi, S., Salaris, M., Castelli, F., 2004, \apj, 612, 168

\bibitem[\protect\citeauthoryear{Pietrinferni et~al.}{2006}]{Pietrinferni06}
Pietrinferni, A., Cassisi, S., Salaris, M., Castelli, F., 2006, \apj, 642, 797

\bibitem[Rayner et al.(2009)]{IRTFII}
Rayner, J.~T., Toomey, D.~W., Onaka, P.~M., Denault, A.~J.,
Stahlberger, W.~E., Vacca, W.~D., Cushing, M.~C., Wang, S., 2003, PASP, 115, 362

\bibitem[\protect\citeauthoryear{Reimers}{1977}]{Reimers}
Reimers, D., 1977, \aap, 57, 395

\bibitem[R\"ock et al.(2015)]{RV:15}
R\"ock, B., Vazdekis, A., Peletier, R.~F., Knapen, J.~H., Falc\'on-Barroso, J., 2015, MNRAS, 449, 2853

\bibitem[R\"ock et al.(2016)]{RV:16}
R\"ock, B., Vazdekis, A., Ricciardelli, E., Peletier, R. F., Knapen, J. H., Falc\'on-Barroso, J., 2016, A\&A, 589, 73

\bibitem[Saglia et al.(2010)]{Saglia:2010}
Saglia, R.~P., et al., 2010, A\&A, 509, A61

\bibitem[\protect\citeauthoryear{Salpeter}{1955}]{Salpeter55}
Salpeter, E.~E., 1955, \apj, 121, 161

\bibitem[\protect\citeauthoryear{S{\'a}nchez-Bl{\'a}zquez et~al.}{2006}]{MILESI}
S{\'a}nchez-Bl{\'a}zquez, P., et~al., 2006, \mnras,  371, 703

\bibitem[Sch\"onebeck et al.(2014)]{SCH:2014}
Sch\"onebeck, F., Puzia, T.~H., Pasquali, A., Grebel, E.~K.,
Kissler-Patig, M., Kuntschner, H., Lyubenova, M., Perina, S., 2014, A\&A, 572, 13 

\bibitem[Silva, Kuntschner, Lyubenova(2008)]{S08}
Silva D.~R., Kuntschner H., Lyubenova, M., 2008, ApJ, 674, 194

\bibitem[Simard et al.(2011)]{Simard:11}
Simard, L., Mendel, J.~T., Patton, D.~R., Ellison, S.~L., McConnachie, A.~W., 2011, ApJS, 196, 11

\bibitem[Smith \& Lucey(2013)]{SmithLucey2013}
Smith, R.~J., Lucey, J.~R., 2013, MNRASL, 434, 1964 

\bibitem[\protect\citeauthoryear{Smith et al.}{2015}]{smith:2015} 
Smith, R.~J., Alton, P., Lucey, J.~R., Conroy, C., Carter, D., 2015, MNRAS, 454, 71 (SAL15)


\bibitem[Sparks, Carollo, Macchetto(1997)]{Sparks:1997}
Sparks, W.~B., Carollo, C.~M., Macchetto, F., 1997, ApJ, 486, 253

\bibitem[Spiniello et al.(2012)]{Spiniello:2012}
Spiniello, C., Trager, S.~C., Koopmans, L.~V.~E., Chen, Y.~P., 2012, ApJ, 753, L32 

\bibitem[Spiniello et al.(2014)]{Spiniello:2014}
Spiniello, C., Trager, S.~C., Koopmans, L.~V.~E., Conroy, C., 2014, MNRAS, 438, 1438 

\bibitem[\protect\citeauthoryear{Spiniello et al.}{2015}]{spiniello:2015} 
Spiniello, C., Trager, S.~C., Koopmans, L.~V.~E., 2015, ApJ, 803, 87

\bibitem[\protect\citeauthoryear{Spinrad \& Taylor}{1971}]{SpinTa:71}
Spinrad H., Taylor B.~J., 1971, ApJS, 22, 445 

\bibitem[\protect\citeauthoryear{Thomas et al.}{2003a}]{TMB:03}
Thomas D., Maraston C., Bender R., 2003, MNRAS, 339, 897

\bibitem[\protect\citeauthoryear{Thomas, Maraston \& Johansson}{2011}]{TMJ11}  
Thomas, D., Maraston, C., Johansson, J., 2011, MNRAS, 412, 2183

\bibitem[\protect\citeauthoryear{Trager et al.}{1998}]{Trager98} 
Trager, S.~C., Worthey, G., Faber, S.~M., Burstein, D., 
Gonz\'alez, J.~J., 1998, ApJS, 116, 1

\bibitem[\protect\citeauthoryear{Treu et al.}{2010}]{Treu:10} 
Treu T., Auger M.~W., Koopmans L.~V.~E., Gavazzi R., Marshall P.~J., 
Bolton A.~S., 2010, ApJ, 709, 1195 

\bibitem[Tripicco \& Bell(1995)]{TripiccoBell:1995} 
Tripicco, M.~J., Bell, R.~A., 1995, AJ, 110, 3035


\bibitem[\protect\citeauthoryear{Tsuji}{1973}]{Tsuji73}
Tsuji T., 1973, A\&A, 23, 411 


\bibitem[\protect\citeauthoryear{Valdes et~al.}{2004}]{Valdes04}
Valdes, F., Gupta, R., Rose, J.~A., Singh, H.~P., Bell, D.~J., 2004, \apjs, 152, 251

\bibitem[\protect\citeauthoryear{van Dokkum \& Conroy}{2010}]{vdC:10}
van Dokkum P.~G., Conroy C., 2010, Nature, 468, 940 

\bibitem[\protect\citeauthoryear{Vazdekis et al.}{1996}]{Vazdekis1996}
Vazdekis, A., Casuso, E., Peletier, R.~F., Beckman, J.~E., 1996, ApJS, 106, 307

\bibitem[Vazdekis et al.(1997)]{Vazdekis1997}
Vazdekis A., Peletier R. F., Beckman J. E., Casuso E., 1997, ApJS, 111, 203

\bibitem[\protect\citeauthoryear{Vazdekis et~al.}{2003}]{CATIV}
Vazdekis, A., Cenarro, A.~J., Gorgas, J., Cardiel, N., 
Peletier, R.~F., 2003, \mnras, 340, 1317

\bibitem[\protect\citeauthoryear{Vazdekis et~al.}{2010}]{MILESIII}
Vazdekis A., S{\'a}nchez-Bl{\'a}zquez P., Falc{\'o}n-Barroso J.,
Cenarro A.~J., Beasley M.~A., Cardiel N., Gorgas J., Peletier R.~F., 2010, \mnras, 404, 1639

\bibitem[\protect\citeauthoryear{Vazdekis et al.}{2012}]{Vazdekis:12}  
Vazdekis, A., Ricciardelli, E., Cenarro, A.~J., Rivero-Gonz\'alez, J.~G.,
D\'iaz-Garc\'a, L.~A., Falc\'on-Barroso, J., 2012, MNRAS, 424, 157 

\bibitem[\protect\citeauthoryear{Vazdekis et al.}{2015}]{Vazdekis:15} 
Vazdekis, A., Coelho, P., Cassisi, S., Ricciardelli, E., 
Falc\'on-Barroso, J., S\'anchez-Bl\'azquez, P., 
La Barbera, F., Beasley, M.~A., Pietrinferni, A., 2015, MNRAS, 449, 1177 (V15)

\bibitem[Ventura et al.(2013)]{Ventura:2013}
Ventura, P., Di Criscienzo, M., Carini, R., D'Antona, F., 2013, MNRAS, 431, 3642

\bibitem[Vernet et al.(2011)]{Vernet:2011} 
Vernet, J., et al., 2011, A\&A, 536, A105

\bibitem[Weidner et al.(2013)]{Weidner2013}
Weidner, C., Ferreras, I., Vazdekis, A., La Barbera, F., 2013, MNRAS, 435, 2274

\bibitem[Worthey, Tang, Baitian(2014)]{Worthey:2014}
Worthey, G., Tang, B., Serven, J., 2014, ApJ, 783, 20W

\bibitem[\protect\citeauthoryear{Zieleniewski et al.}{2015}]{Ziel:15}
Zieleniewski S., Houghton R.~C.~W., Thatte N., Davies R.~L., 2015,
MNRAS, 452, 597

\end{thebibliography}







\bsp	
\label{lastpage}
\end{document}